\documentclass[11pt,preprint]{aastex}

\shorttitle{Kinematics of Galactic thick disk}
\shortauthors{Moni Bidin et~al.}

\begin{document}

\title{Kinematical and chemical vertical structure of the Galactic thick
disk\thanks{Based on observations collected at the European Organization for Astronomical Research in the Southern
Hemisphere, Chile (proposal IDs 075.B-0459(A), 077.B-0348(A))}~$^{,}$\thanks{This paper includes data gathered with the
6.5-meter Magellan and the duPont Telescopes, located at Las Campanas Observatory, Chile.} \\ I. Thick disk kinematics}

\author{C. Moni Bidin}
\affil{Universidad de Concepci\'on, Departamento de Astronom\'ia, Casilla 160-C, Concepci\'on, Chile}
\email{cmbidin@astro-udec.cl}
\author{G. Carraro\altaffilmark{3}}
\affil{European Southern Observatory, Alonso de Cordova 3107, Vitacura, Santiago, Chile}
\altaffiltext{3}{Universit\'a di Padova, Dipartimento di Astronomia, Vicolo Osservatorio 3, I-35122, Padova, Italia}
\author{R. A. M\'endez}
\affil{Universidad de Chile, Departamento de Astronom\'ia, Casilla 36-D, Santiago, Chile}

\begin{abstract}
The variation of the kinematical properties of the Galactic thick disk with Galactic height $Z$ are studied by means
of 412 red giants observed in the direction of the south Galactic pole up to 4.5~kpc from the plane. We confirm the
non-null mean radial motion toward the Galactic anticenter found by other authors, but we find that it changes sign at
$\vert Z \vert$=3~kpc, and the proposed inward motion of the LSR alone cannot explain these observations. The
rotational velocity decreases with $\vert Z \vert$ by $-30$~km~s$^{-1}$~kpc$^{-1}$, but the data are better
represented by a power-law with index 1.25, similar to that proposed from the analysis of SDSS data. All the velocity
dispersions increase with $\vert Z \vert$, but the vertical gradients are small. The dispersions grow proportionally,
with no significant variation of the anisotropy. The ratio $\sigma_\mathrm{U}$/$\sigma_\mathrm{W}$=2 suggests that the
thick disk could have formed from a low-latitude merging event. The vertex deviation increases with Galactic height,
reaching $\sim$20 degrees at $\vert Z \vert$=3.5~kpc. The tilt angle also increases, and the orientation of the
ellipsoid in the radial-vertical plane is constantly intermediate between the alignment with the cylindrical and the
spherical coordinate systems. The tilt angle at $\vert Z \vert$=2~kpc coincides with the expectations of MOND, but an
extension of the calculations to higher $\vert Z \vert$ is required to perform a conclusive test. Finally, between 2.5
and 3.5~kpc we detect deviations from the linear trend of many kinematical quantities, suggesting that some kinematical
substructure could be present.
\end{abstract}



\section{INTRODUCTION}
\label{s_intro}

The formation mechanism of the thick disk is one of the most noticeable grey points of our current understanding of the
Galactic formation and evolution process. This uncertainty is particularly unfortunate because, being the thick disk a
ubiquitous feature among spiral galaxies \citep{Dalcanton02,Seth05}, its formation must be a common stage in the early
evolution of late-type galaxies. During the nearly three decades since its discovery \citep{Gilmore83}, many models have
been proposed. The merging scenario, in which the early merging of a small satellite galaxy heats a primordial Galactic
disk producing an old, thick, and kinematically hot disk population \citep{Quinn93,Walker96}, has been the most favored
model in the last decade, following the evidence that the thin and thick disks are chemically distinct
\citep{Fuhrmann98,Gratton01}. Nevertheless, even this scenario is not free of problems
\citep[see, for example,][]{Bournaud09}, and alternative models have recently been drawn to attention
\citep[e.g.,][]{Bournaud09,Schonrich09,Roskar08,Assmann11}.

It is clear that the general properties of the Galactic thick disk, such as its mean metallicity or mean kinematics,
are not enough to distinguish between the models proposed for its formation. Moreover, the merging scenario has many
free parameters, such as the mass of the merging satellite and the inclination of its orbit with respect to the
Galactic disk, and the observations must constrain them if the quality of the model is to be finally assessed. In the
last few years, theoretical simulations have started to cast predictions of the stellar distribution of stars in the
multi-dimensional spatial-kinematical-chemical space \citep[e.g.,][]{Loebman11,Kobayashi11}. For example,
\citet{Gomez11} have shown that, within the merging scenario, the time of impact, and the mass and orbit of the
infalling satellite can be deduced from the distribution of the expected kinematical clumps in the $U$--$V$ plane,
while \citet{Villalobos08,Villalobos09}, \citet{Villalobos10}, and \citet{Purcell09} find that the
$\sigma_\mathrm{U}$/$\sigma_\mathrm{W}$ ratio and its variation with Galactocentric distance are excellent indicators of
the inclination angle of the merging event. \citet{Villalobos08} and \citet{Bekki11} have also shown that the
observed decrease of rotation velocity with distance from the plane points to a low-latitude merging. At the same time,
the observations are gathering information about the spatial variations of the chemical composition and velocity
distribution \citep[e.g.][]{Ivezic08,Fuchs09,Bond10,Dinescu11}. Detailed results of this kind are strongly needed
because, through comparison with the expectations from the different models, they can be key to solve the puzzle of the
Galactic thick disk formation.

We are undertaking an extensive survey aimed to reveal the kinematical and chemical vertical structure of the
Galactic thick disk \citep{Carraro05}. Preliminary kinematical results were presented by \citet{Moni08} and
\citet{Moni09}, while the sample was also used to investigate the Galactic dark disk \citep{Moni10}, and Lithium-rich
stars in the Galactic thick disk \citep{Monaco11}. In this paper, we focus on the kinematical results, studying the
trend of kinematical quantities as a function of distance from the Galactic plane. In later papers of this series,
the collected spectra will be used to measure the metallicity and chemical abundances of the sample, to study the
variation of the thick disk chemistry with Galactic height.


\section{THE SAMPLE}
\label{s_sample}

\begin{figure}
\epsscale{1.}
\plotone{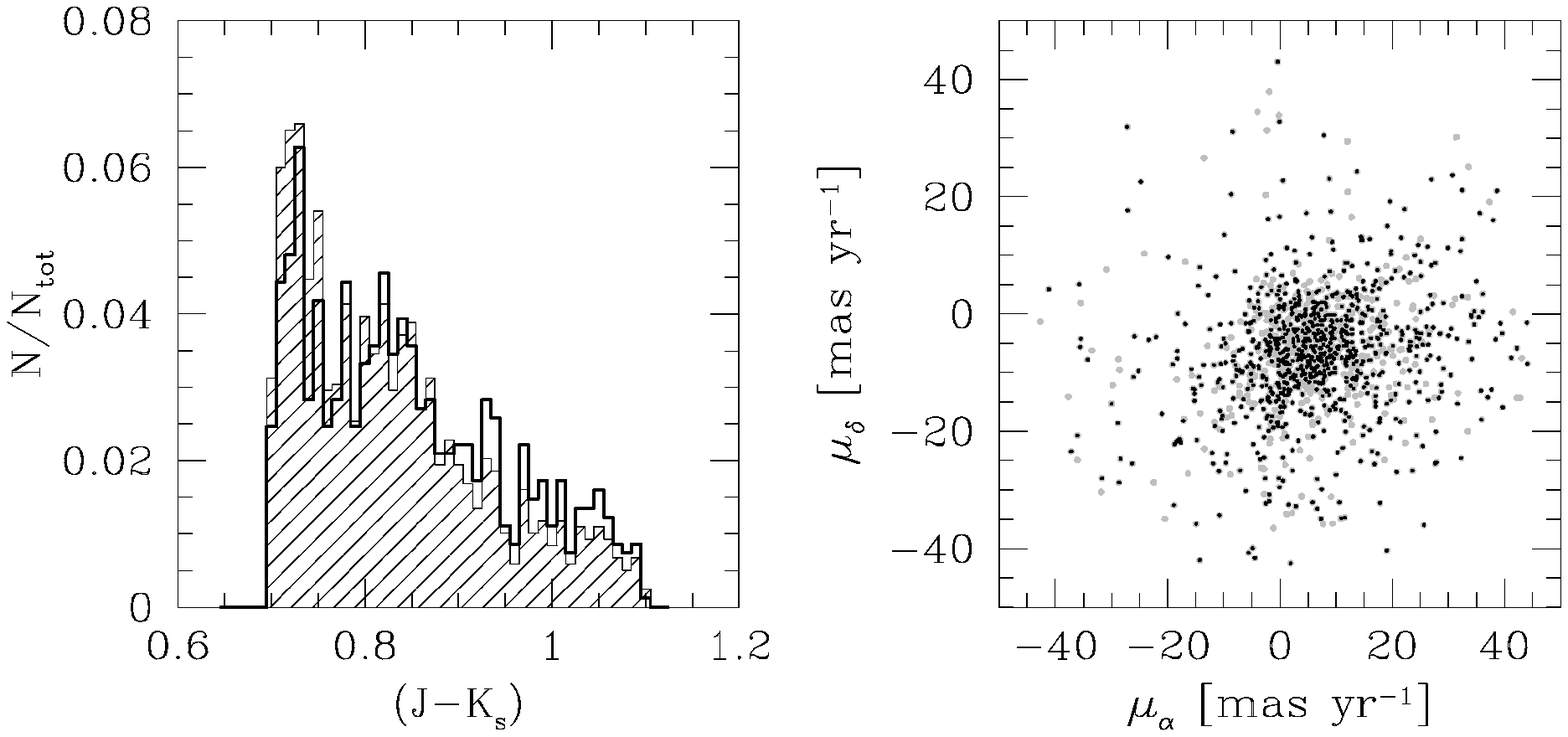}
\caption{Left panel: color distribution of the spectroscopically observed stars (thick line) and of the complete sample
(shaded histogram); Right panel: proper motion distribution of the spectroscopically observed stars (black dots) and
of the complete sample (grey dots).\label{f_sample}}
\end{figure}

Our investigation is based on the sample of $\sim$1200 red giants defined by \citet{Girard06}, vertically distributed
with respect to the Galactic plane in a cone of fifteen degrees radius centered on the South Galactic Pole. The sample
covers a large range of Galactic heights, from the plane to $\vert Z \vert\sim$5~kpc, and it is volume-complete up to
$\vert Z\vert$=3~kpc. All the objects have 2MASS photometry \citep{Skrutskie06}, and absolute proper motions from the
SPM3 catalog \citep{Girard04}. The sample was defined by the color cut 0.7$\leq(J-K_\mathrm{s})\leq$1.1, to isolate
intermediate-metallicity stars. Main Sequence (MS) dwarfs were excluded both by a sloped cut at fainter magnitudes,
parallel to the MS, which excludes all but the nearest (d$\leq$63~pc) dwarfs, and by conservative kinematical criteria
imposing a stellar velocity lower than the local escape velocity \citep[see][for more details]{Girard06}.

\begin{table*}
\begin{center}
\caption{Log of the spectroscopic observations \label{t_obs}}
\begin{tabular}{l l c l c}
\tableline\tableline
Run & Date & N. nights & Instrument & N. stars \\
\tableline
Coralie1 & 2005, September  & 4 & Coralie@Euler     & 39  \\
FEROS1   & 2005, September  & 6 & FEROS@2.2         & 183 \\
duPont1  & 2005, October    & 6 & Echelle@duPont    & 168 \\
Coralie2 & 2005, October    & 3 & Coralie@Euler     & 25  \\
FEROS2   & 2006, August     & 6 & FEROS@2.2         & 161 \\
duPont2  & 2006, September  & 6 & Echelle@duPont    & 119 \\
MIKE1    & 2006, Aug.--Nov. & 5x0.5 & MIKE@Magellan & 77  \\
MIKE2    & 2007, October    & 2 & MIKE@Magellan     & 52  \\
\tableline
\end{tabular}
\end{center}
\end{table*}

We collected high-resolution Echelle spectra for two-thirds of the Girard et~al.'s sample (824 stars). All the stars in
the brighter half of the sample were spectroscopically observed, while the completeness decreases with magnitude for
fainter objects. We found a high contamination ($\sim$35\%) by nearby dwarf in the faintest end of the distribution,
hence stars with $K_\mathrm{s}\geq9.5$ were given lower priority after the first observations. The distribution of
proper motions and colors of the observed sub-sample is shown in Figure~\ref{f_sample}. The comparison with the complete
sample reveals that no selection effect was introduced: the slight overabundance of red stars in the observed sample
is due to the higher completeness at brighter magnitudes, where stars are on average redder.

As discussed later, we will restrict our investigation to stars with Galactic height $\vert Z\vert\geq$1.3~kpc, to avoid a
strong thin disk contamination. This reduced the sample under study to 514 stars. We visually inspected all the
spectra, identifying 46 probable dwarf stars. As expected, they show on average low radial velocities but high U and V
components, and only one is found at $K_\mathrm{s}\leq$9. We also found 22 stars with [Fe/H]$\leq-$1.5, as deduced
from comparisons with synthetic spectra. They were considered probable halo contaminants and were excluded
from further analysis. We note that the metal-poor tail of the thick disk extends to much lower metallicities
\citep{Beers95}, but below this limit thick disk stars are outnumbered by halo members by a factor of nearly ten
\citep{Chiba00}. Finally, we also excluded from analysis 34 stars that, in the Galactic cylindrical coordinate system, had
velocity components (defined in Section~\ref{s_disp}) outside the range $\vert$U$\vert\leq$300~km s$^{-1}$,
$-500\leq$V$\leq$300~km s$^{-1}$, and $\vert$W$\vert\leq$150~km s$^{-1}$. The cut in $W$ was aimed to remove the residual
halo contamination, while the other components were restricted to exclude stars with wrong distance or bad proper motion.
The cut in $V$ was offset toward negative values to take into account the vertical shear (see Section~\ref{s_resmean}).
Our final sample thus comprised 412 stars.

\subsection{Distances}
\label{s_distance}

The interstellar reddening E($B-V$) was derived for each target from the \citet{Schlegel98} maps, and the $K_\mathrm{s}$
magnitude and ($J-K_\mathrm{s}$) color were de-reddened by means of the transformations of \citet{Cardelli89}. The
distance to each star was then estimated through a color-absolute magnitude relation calibrated on 47\,Tucanae, a disk
globular cluster \citep{Zinn85} whose stellar population is very similar to the Galactic thick disk both in age and
metallicity \citep{Wyse05}. The fit of the cluster red giant branch yields the relation (L. Miller 2008, priv. comm.):
\begin{equation}
K_\mathrm{s}= -7.886\cdot (J-K_\mathrm{s}) + 16.302,
\label{eq_dist}
\end{equation}
which were translated into absolute magnitude and de-reddened color assuming $(m-M)_V$=13.50$\pm$0.08 and
E($B-V$)=0.024$\pm$0.004 for the cluster distance modulus and reddening, respectively \citep{Gratton03}.

Inspecting the 2MASS photometric data of 47\,Tuc used to derive Equation~(\ref{eq_dist}), we found that the
uncertainty on $M_K$ is of the order of $\sim$0.2 magnitudes. This is only a marginal contribution to the final error
in distance, because the main source of uncertainty is the relatively wide range of age and metallicity covered by thick
disk stars. Indeed, the 2MASS photometric errors have only negligible impact, because they do not exceed
0.03~magnitudes for our fainter targets ($K_\mathrm{s}\sim$10.6). We estimated the effect of the age and metallicity
distribution on the derived absolute magnitudes by means of Yale-Yonsei isochrones \citep{Yi03}. We assumed a scatter
of 0.3~dex in metallicity, that should include the bulk of thick disk stars \citep{Carney89} when excluding the
scarcely-populated low- and high- metallicity tails \citep{Beers95,Bensby07}, and a scatter of 2~Gyr in age
\citep{Bensby03,Feltzing03,Reddy06}. They were considered uncorrelated, because the age-metallicity relation for
the thick disk is very weak \citep{Bensby07}. We finally estimated the error on distance to be $\sim$20\%, quadratically
summing all the relevant contributions.

Thin disk stars do not follow the age and metallicity distribution assumed to derive Equation~(\ref{eq_dist}), and their
distances should be systematically biased. Indeed, younger, more metal-rich red giants are intrinsically fainter than
our estimate. Comparing the absolute magnitudes calculated from Equation~(\ref{eq_dist}) with Yale-Yonsei isochrones
between 2 and 8~Gyr, and metallicity following the age-metallicity relation of \citet{Haywood01} and
\citet{RochaPinto00}, we found that the distance of thin disk stars would be overestimated by 10--20\%. This
systematic error is small, but it has a relevant consequence on the contamination of the sample, which is artificially
stretched to larger heights from the plane.

\subsection{Thin disk contamination}
\label{s_TnDcont}

\begin{figure}
\epsscale{1.}
\plotone{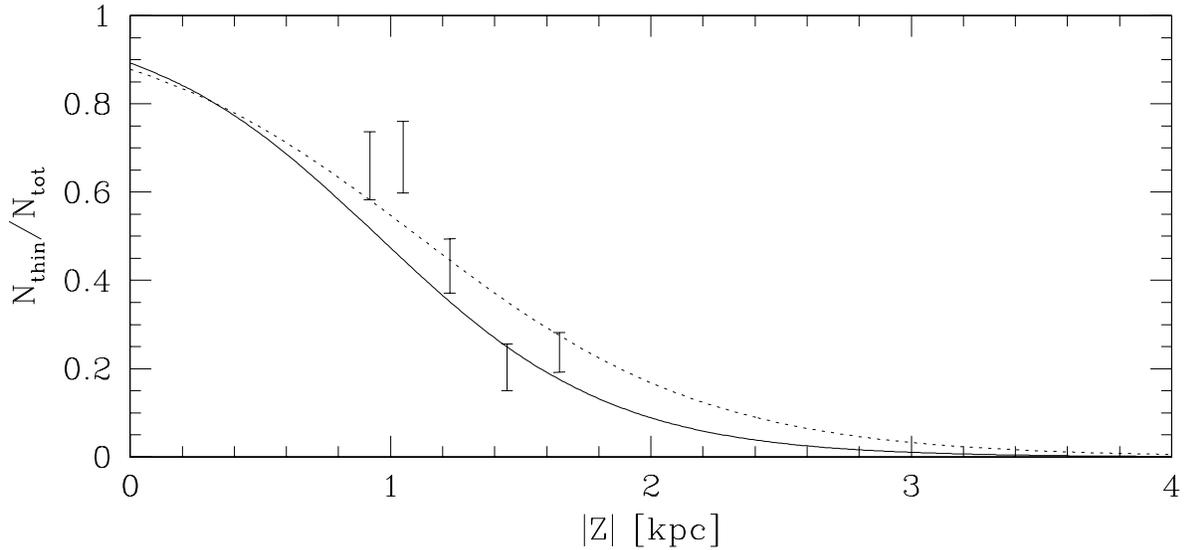}
\caption{Estimated fraction of thin disk stars in the sample. The curves indicate the expectation of the model described
in the text, when a distance overestimate of 15\% for thin disk is considered (dashed curve) or not (thick curve). The
errorbars show the results of our empirical estimate. \label{f_contam}}
\end{figure}

Our sample is contaminated by the thin disk, because the targets were selected through a color cut that efficiently
eliminates low-metallicity stars, but which excludes only a small fraction of metal-rich objects, as it can be
deduced from Figure~1 of \citet{Girard06}. In Figure~\ref{f_contam} we plot an estimate of the fraction of thin disk stars
in the sample, assuming 300 and 900~pc as thin and thick disk scale height, respectively, and a thick disk local
normalization of 12\% \citep{Juric08}. We also calculated the contamination in the case of a distance overestimate of 15\%
for thin disk objects. The curves can be considered an upper limit, because the color cut, unaccounted for in the
calculation, could have removed part of the contamination. In the same figure we indicate the results of a rough empirical
estimate, obtained fitting a double Gaussian to the distribution of the vertical velocity component ($W$) in five bins of
50~stars. The dispersions of the two Gaussian components were kept fixed, and the only free parameter of the fit was the
number of thin disk stars in the bin. The error bars show the results when varying $\sigma_\mathrm{W}$ in the
range 14-16 and 35-40~km~s$^{-1}$ for the thin and the thick disk, respectively. This test was performed only for
$\vert Z\vert\approx$1-2~kpc, where both populations contribute by more than 20\% of the objects, i.e. 10 stars in each bin.
The empirical errorbars agree well with the curve obtained when the distance bias is considered, although the observed
contamination fall-off with $\vert Z \vert$ seems steeper than the modeled one.

In conclusion, our sample is contaminated by thin disk stars, whose distances are overestimated. This affects
even the estimate of their kinematics, which is also biased. Therefore, we will not analyze the results for the thin disk,
and we will limit our study to $\vert Z\vert\geq$1.3~kpc, to avoid the strong contamination ($\geq$50\%) of the nearest
subsample.

\subsection{Halo contamination}
\label{s_Hcont}

\citet{Girard06} estimated that the halo contamination in the sample should be about 8\%, hence we would expect 41 halo
members among the 514 targets with $\vert Z\vert\geq$1.3~kpc. The cut in $W$ removed 10 probable halo contaminants and,
according to the statistics of \citet{Chiba00}, 18 of the excluded low-metallicity targets are expected to be halo stars. The
cuts in $U$ and $V$ could also have removed some halo objects, but the effect of this cut is harder to
quantify, because an unknown fraction of the outliers are probably objects with problematic proper motion or wrong distance.
The residual halo contamination in our sample of 412 targets should therefore be lower than $\sim$3\% (13 stars), and it
can be safely neglected.


\section{OBSERVATIONS AND DATA REDUCTION}
\label{s_reduction}

The spectra were collected during six observing runs between 2005 and 2007, at La Silla and Las Campanas observatories.
The details of the observations and data reduction were presented in \citet{Moni09}, and we will outline here only
the most relevant points. Four different telescopes were used, because the stars span a wide range in magnitude
($V$=5--16), but the instruments had similar characteristics. The basic information of the runs is given in
Table~\ref{t_obs}. The spectra covered the whole visual range from the atmospheric cutoff on the blue side to $\sim$9000~\AA,
except for Coralie data which only reached 6800~\AA\ on the red edge. We did not reduce the spectra collected
with the MIKE blue arm because of too low stellar flux, and MIKE data were thus limited to 4850~\AA\ on the blue end. The
spectral resolution varied between 32\,000 and 50\,000, depending on the spectrograph. In particular, the 0$\farcs$7 slit
was always used for MIKE runs (R=32\,000), while at duPont both the 0$\farcs$75 and 1$\arcsec$ slits were used (R=40\,000
and 30\,000, respectively), depending on weather conditions. During each run we secured the spectra of up to
seventeen bright stars with accurate parameters from the literature (radial and rotational velocities,
temperature, gravity, metallicity). Exposure times were chosen according to weather conditions, to reach
S/N=70--100 for all the targets. Comparison lamp arcs were acquired only during daytime calibrations for the fiber
spectrographs Coralie and FEROS. The second fiber of these two instruments was allocated to the sky background, because
the spectra were usually collected next to full moon. Lamp arcs were collected at intervals of about 2 hours during duPont
and MIKE runs, and each spectrum was calibrated with the average of the two lamps temporally enclosing it, linearly
weighted by the temporal distance between science and calibration frames.

Spectra were reduced with standard IRAF\footnote{\small{IRAF is distributed by the National Optical Astronomy
Observatories, which are operated by the Association of Universities for Research in Astronomy, Inc., under
cooperative agreement with the National Science Foundation.}} tasks, and we kept the procedures as much as possible
identical for all the data. We a posteriori verified that the reduced FEROS and Coralie
spectra were of the same quality as the products of their dedicated pipelines. We analyzed the dark frames collected
for all the runs, but we always found a negligible instrumental dark current and no clear 2D pattern, hence no dark
correction was applied. The wavelength calibration lamp spectra were extracted in the same position on the CCD as
science targets, to avoid systematics introduced by the curvature of the orders. The spectrum of a bright,
well-exposed star was used to trace the orders on the CCD in each observing night, allowing only for rigid shifts
among the frames. Then, the spectra were extracted with an optimum extraction algorithm \citep{Horne86}, normalized,
and finally merged.


\section{MEASUREMENTS}
\label{s_meas}

\subsection{Radial velocities}
\label{s_rv}

The radial velocity (RV) for all the targets was measured with a cross-correlation (CC) technique \citep{Tonry79}
as implemented in the IRAF {\it fxcor} task. The procedure was described in detail in \citet{Moni09}, and we give only
a brief summary here. The spectrum of each star was cross-correlated with three template stars observed in the
same run, namely HD\,180540, HD\,223559, and HD\,213893 (this last replaced by HD\,224834 for Coralie data),
encompassing the temperature range of the targets. Their RV was fixed by a CC with a synthetic spectrum generated
with the SPECTRUM\footnote{http://www.phys.appstate.edu/spectrum/spectrum.html} code \citep{Gray94}, because we
found poor agreement between the available literature sources. The analysis of the seventeen standard stars
acquired during observations, and of the solar spectra collected each night, revealed that the RV zero-point thus
defined was biased by 0.3-0.7~km~s$^{-1}$, depending on the instrument, and this offset was removed.

The three measurements were averaged, although they never differed by more than 0.2~km~s$^{-1}$. The CC was
restricted to the range 4600-6800~\AA\ (5000-6800~\AA\ for MIKE data), to avoid the poor-quality blue section and
the telluric bands at longer wavelengths. All RVs were corrected to heliocentric velocities, then the position of
the telluric band at 6875~\AA\ was used to correct the RVs for instrumental effects, mainly caused by a
displacement on CCD between the lamp and science spectra, and an off-center position of the star inside the slit
\citep[see, for example, the analysis of][]{Moni06}. Corrections of up to 2~km~s$^{-1}$ were applied, but with
little scatter ($\sim$0.5~km~s$^{-1}$) within each observing night.

The final RV errors were estimated as the quadratic sum of the five relevant uncertainties introduced in the
procedure: the CC and wavelength calibration error, the uncertainty on the zero-point and its offset, and the error on
the estimate of the correction for instrumental effects. The resulting errors were typically in the range
0.4-0.7~km~s$^{-1}$. The final RVs of all the 824 stars will be made available at the CDS
website\footnote{http://cdsweb.u-strasbg.fr/}.

\begin{table}
\begin{center}
\caption{Mean RV difference between this work and the quoted reference for the stars in common. \label{t_lit}}
\begin{tabular}{l c c}
\tableline\tableline
Reference & N. stars & $\overline{\Delta_\mathrm{RV}}$ \\
 &  & km s$^{-1}$ \\
\tableline
\citet{Flynn93}    & 145 & 0.8$\pm$2.5 \\
\citet{Zwitter08}  & 9   & $-1.2\pm$1.5 \\
\citet{Beers95}    & 9   & 0.4$\pm$1.8 \\
\citet{Majewski04} & 8   & $-4.7\pm$2.1 \\
\citet{Duflot95}   & 6   & $-1.1\pm$3.8 \\
\citet{Jones72}    & 6   & $-0.7\pm$0.9 \\
\citet{Evans70}    & 6   & 1.8$\pm$4.2 \\
\tableline
\end{tabular}
\end{center}
\end{table}

We found 211 previous RV measurements for the stars in our sample. Our results agree excellently with literature
sources: the mean difference (in the sense ours$-$literature) is 0.4$\pm$2.7~km~s$^{-1}$, where the uncertainty is
the rms of the differences. The comparison with the works that share with us more than five stars in common is
given in Table~\ref{t_lit}. The mean RV difference is always of the order of 1~km~s$^{-1}$, except for
\citet{Majewski04}, whose RVs are higher than ours by a non-negligible amount ($\sim$5~km~s$^{-1}$). However, the
number of stars in common is too small to conclude that this offset is significant.

\subsection{Galactic velocities}
\label{s_disp}

The proper motion, radial velocity, and distance of each target were combined to derive its ($U,V,W$)
velocity components in the Galactic cylindrical reference frame, where $U$ is positive toward the Galactic center,
$V$ is directed in the sense of Galactic rotation, and $W$ points toward the North Galactic Pole. The error on
these velocities was derived propagating the uncertainty on the observed quantities. We assumed a proper motion
error of 3~mas~yr$^{-1}$ for all the stars, as this value is more realistic than the uncertainties quoted in the
SPM3 catalog (T. Girard, 2009, priv. comm.; see also \citealt{Girard06}). The velocities were corrected subtracting
the solar peculiar motion ($U_\odot,V_\odot,W_\odot$)=(10.0, 5.1, 7.2)~km~s$^{-1}$ \citep{Dehnen98}.
\citet{Schonrich10} recently proposed the revised values ($U_\odot,V_\odot,W_\odot$)=(11.0, 12.2, 7.3)~km~s$^{-1}$,
but we preferred to adopt the older ones, for sake of continuity with previous works. In any case, the definition
of the solar motion does not affect the velocity dispersions nor the off-diagonal terms of the dispersion matrix
(Equation~\ref{e_cross}), while the effects on the mean values are discussed in Section~\ref{s_resmean}.

The sample was
then divided into several bins, in which we calculated the mean velocities, the dispersions
($\sigma_\mathrm{U},\sigma_\mathrm{V},\sigma_\mathrm{W}$), and the non-diagonal terms of the dispersion matrix
\begin{equation}
\sigma^2_{\alpha\beta}=\frac{1}{(N-1)}\Sigma_i (v_{\alpha,i}-\overline{v_{\alpha}})(v_{\beta,i}-\overline{v_{\beta}}),
\label{e_cross}
\end{equation}
where the sum is extended to all the stars in the bin, and $\alpha,\beta=U,V,W$. The results are given in
Table~\ref{t_res}. The bins were defined by the requirement that their centers were equally spaced by 0.1~kpc from
$\vert Z \vert$=1.5 to 4.5~kpc, to uniformly sample the variation with $\vert Z \vert$ of the kinematical
quantities, while the width was imposed by the fixed number of stars per bin (see below). This implied a partial
overlap of the bins, increasing with distance from the plane due to the decreasing density of observed stars. However,
\citet{Moni10} have shown that the binning definition does not alter the results, and in fact our results are very
similar to that work, despite the very different binning schemes. The bins with $\overline{\vert Z\vert}\geq$2.4~kpc
comprised 50~stars each, while at lower heights, where the number of observed stars is larger, the bin size
was increased to 80~targets for 2.1$\leq\overline{\vert Z\vert}\leq$2.4~kpc, and 100 targets for
$\overline{\vert Z\vert}\leq$2.1~kpc. We thus ensured a good statistic of thick disk stars in bins contaminated by the
thin disk.

\begin{figure}
\epsscale{1.}
\plotone{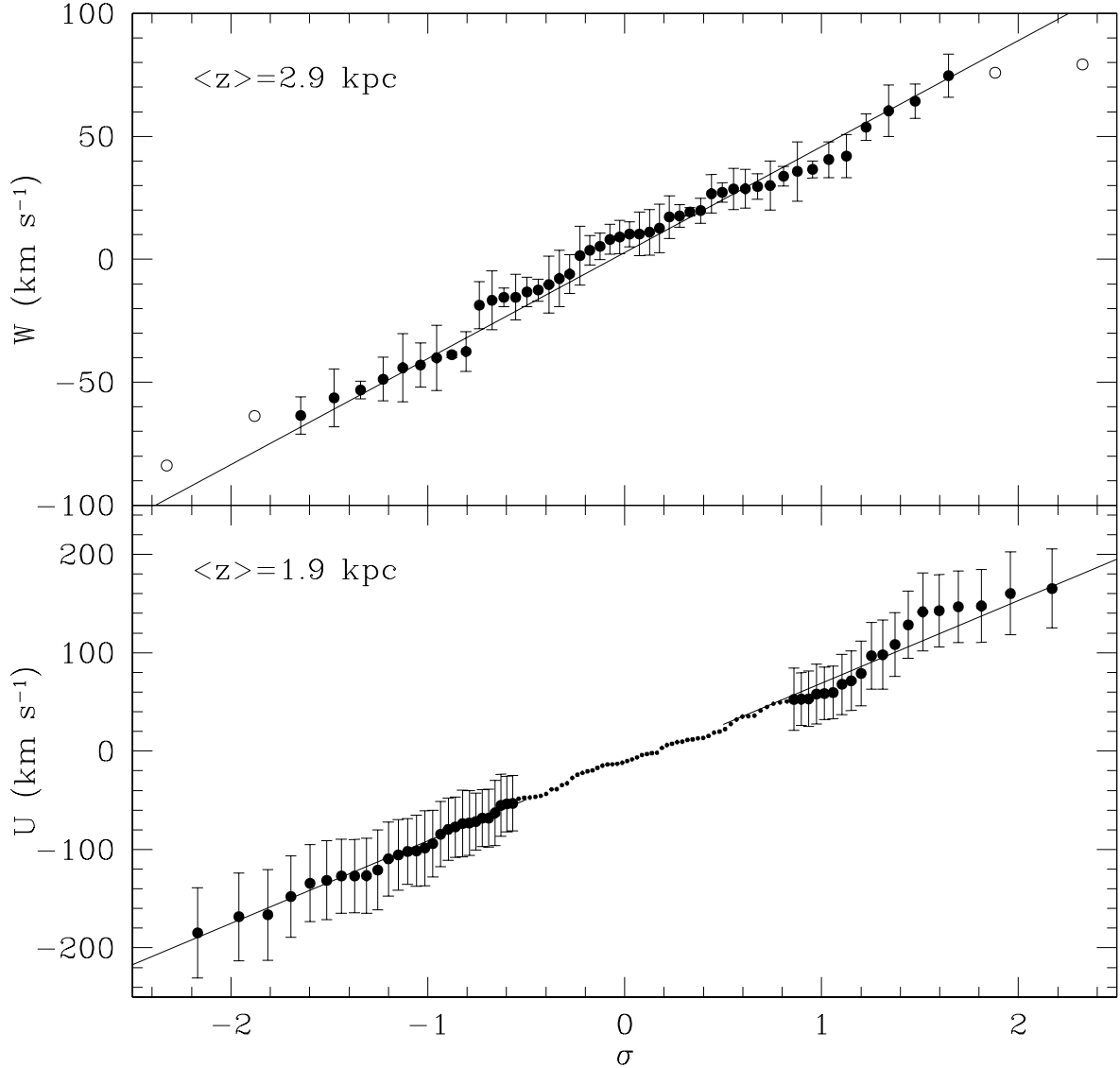}
\caption{
Examples of probability plots used to derive the mean value and the dispersion of the velocity components. The
velocity is plotted against the $\sigma$-value, assigned to each datum as described in the text. The mean value and
the dispersion are given, respectively, by the intercept and the slope of the linear fit to the points in the plot.
Upper panel: bin of 50 stars with $\vert Z\vert\geq$2.5~kpc. The line shows the fit, and the empty dots indicate the
data excluded from the fit. Lower panel: bin of 100 stars in the region contaminated by thin disk stars
($\vert Z\vert\leq$2.5~kpc). Only the wings of the distribution were fitted, and the data not used in the procedure
are shown as small dots.\label{f_probplot}}
\end{figure}

In each bin, the mean value and dispersion of each velocity component were measured by means of the analysis of
the corresponding probability plot \citep{Hamaker78,Lutz92}. This powerful tool is highly insensitive to outliers,
and it can be employed even in case of poorly populated bins. The data are first ordered with increasing value of the
velocity. Each point $i$ is then assigned a value $\sigma$, equal to the deviation from the mean expected for the
$i^{th}$ point of a normal distribution, in units of the standard deviation. When the underlying distribution is
Gaussian, the data follow a linear relation in the $\sigma$-velocity plane, whose intercept and slope are given by
the mean value and the dispersion, respectively. The probability plots were fitted with a straight line, weighting
each data point by its error, two examples are shown in Figure~\ref{f_probplot}. The intrinsic velocity
dispersions were then obtained quadratically subtracting the mean velocity error in the bin. The wings of the
distribution showing any deviation from linearity were excluded from the fit, suspected to be affected by objects
with wrong distance or problematic proper motion, or by residual halo members.

The analysis of artificial data samples, generated through Monte-Carlo simulations and analyzed as the real
data, indicated that the velocity dispersion is underestimated by 1-2~km~s$^{-1}$ when the thin disk
contamination approaches 10\%. The contamination was therefore neglected for $\overline{\vert Z\vert}\geq$2.5~kpc,
where it is expected to be lower than 5\% (i.e. two stars per bin, see Figure~\ref{f_contam}). Following
\citet{Bochanski07}, in the presence of a non-negligible thin disk contamination
($\overline{\vert Z\vert}\leq$2.5~kpc) we
derived the mean velocity and the dispersion of thick disk stars fitting only the wings of the probability plot,
and only the negative one for $V$. The cut excluded the ranges $\vert$U$\vert\leq$50~km~s$^{-1}$,
$\vert$W$\vert\leq$30~km~s$^{-1}$, and $V\geq -$60~km~s$^{-1}$, approximatively 1.5~times the expected thin disk
dispersion, thus removing about 90\% of the contaminants. The two wings of the probability plot should have the
same slope but a different intercept, and the mean velocity was obtained averaging the two intercepts obtained
from the fit. Although the mean values thus derived agree well with the trends observed in the
contamination-free bins at higher $\vert Z \vert$, we regard them as less reliable.

The formal errors of the least-square procedure used to obtain the intercept and the slope in the probability
plots mainly indicate the deviations from Gaussianity of the observed distribution. Hence, they are not a good
estimate of the real uncertainties. The errors were therefore derived by means of Monte-Carlo simulations.
For each bin, we generated one thousand artificial samples of 50 stars, changing the proper motion, distance, and
RV of each star assuming a Gaussian random noise with dispersion equal to the estimated errors (3~mas~yr$^{-1}$,
20\%, and 0.4-0.7~km~s$^{-1}$, respectively). The measurements were repeated in each artificial bin as done for
the real data, and the dispersion of these artificial measurements was assumed as an estimate of the observational
errors. The final errors are given in Table~\ref{t_res}.

The non-diagonal terms of the dispersion matrix, $\sigma^2_\mathrm{UV}$, $\sigma^2_\mathrm{UW}$, and
$\sigma^2_\mathrm{VW}$, were calculated by means of Equation~\ref{e_cross}. The errors were estimated from
Monte-Carlo simulations, as done for the velocity dispersions. For each bin, one thousand artificial data sets of
50~values were simulated, drawn from a distribution characterized by the same dispersion matrix as the real data.
 In each simulation, a Gaussian random noise with a dispersion equal to the observational errors was added,
and the non-diagonal terms were estimated by means of Equation~\ref{e_cross}. The errors were defined by the
scatter of these simulated measurements. With the same simulations we also evaluated the  systematic error
introduced by the observational uncertainties \citep[e.g.][]{Siebert08}, comparing the results when the
random noise was considered or not. We found that the expected systematic thus introduced is negligible, being
about one order of magnitude lower than the uncertainties on the measured values. Therefore, we did not correct
the observed non-diagonal terms for observational errors.

\begin{table*}
\caption{Measured kinematical quantities and associated uncertainties in each bin.\label{t_res}}
\begin{tabular}{c r r r r r r r r r}
\tableline
\tableline
$\vert Z \vert$ & $\overline{U}$ & $\overline{V}$ & $\overline{W}$ & $\sigma_\mathrm{U}$ & $\sigma_\mathrm{V}$ & $\sigma_\mathrm{W}$ & 
$\sigma^2_\mathrm{UW}$ & $\sigma^2_\mathrm{UV}$ & $\sigma^2_\mathrm{VW}$ \\
kpc & km~s$^{-1}$ & km~s$^{-1}$ & km~s$^{-1}$ & km~s$^{-1}$ & km~s$^{-1}$ & km~s$^{-1}$ & km$^2$~s$^{-2}$ & km$^2$~s$^{-2}$ & km$^2$~s$^{-2}$ \\
\tableline
1.5 & $-$31$\pm$4 & $-$46$\pm$4   & 17$\pm$1    & 81$\pm$5  & 57$\pm$5  & 38$\pm$1 & 570$\pm$310  & 400$\pm$470     & 10$\pm$210       \\
1.6 & $-$20$\pm$4 & $-$55$\pm$4   & 5$\pm$1     & 77$\pm$5  & 59$\pm$5  & 37$\pm$1 & 840$\pm$290  & $-$180$\pm$460  & $-$20$\pm$220   \\
1.7 & $-$28$\pm$4 & $-$61$\pm$5   & 3$\pm$1     & 79$\pm$5  & 64$\pm$5  & 38$\pm$1 & 930$\pm$310  & $-$130$\pm$510  & $-$40$\pm$250   \\
1.8 & $-$24$\pm$4 & $-$69$\pm$5   & 8$\pm$1     & 79$\pm$5  & 61$\pm$6  & 40$\pm$1 & 810$\pm$330  & $-$760$\pm$500  & $-$80$\pm$240   \\
1.9 & $-$16$\pm$4 & $-$77$\pm$5   & 3$\pm$1     & 78$\pm$5  & 60$\pm$5  & 40$\pm$1 & 850$\pm$330  & $-$1060$\pm$500 & $-$120$\pm$240  \\
2.0 & $-$10$\pm$5 & $-$85$\pm$4   & 4$\pm$1     & 83$\pm$6  & 55$\pm$6  & 39$\pm$1 & 940$\pm$340  & 30$\pm$460      & $-$90$\pm$210   \\
2.1 & $-$22$\pm$5 & $-$78$\pm$5   & 3$\pm$1     & 80$\pm$6  & 58$\pm$6  & 38$\pm$1 & 1060$\pm$320 & 470$\pm$470     & $-$130$\pm$220  \\
2.2 & $-$16$\pm$5 & $-$76$\pm$7   & $-$13$\pm$1 & 77$\pm$6  & 63$\pm$7  & 42$\pm$1 & 690$\pm$330  & 0$\pm$550       & 230$\pm$260     \\
2.3 & $-$1$\pm$6  & $-$81$\pm$5   & $-$9$\pm$1  & 81$\pm$7  & 58$\pm$6  & 40$\pm$1 & 750$\pm$340  & $-$500$\pm$470  & 10$\pm$230       \\
2.4 & $-$12$\pm$5 & $-$85$\pm$6   & $-$12$\pm$1 & 80$\pm$6  & 59$\pm$6  & 40$\pm$1 & 410$\pm$330  & $-$710$\pm$480  & 180$\pm$240     \\
2.5 & $-$18$\pm$6 & $-$85$\pm$6   & $-$2$\pm$1  & 78$\pm$7  & 63$\pm$7  & 42$\pm$1 & 630$\pm$330  & $-$1870$\pm$600 & $-$40$\pm$270   \\
2.6 & $-$25$\pm$6 & $-$90$\pm$6   & 1$\pm$1     & 71$\pm$7  & 66$\pm$7  & 42$\pm$1 & 530$\pm$310  & $-$2390$\pm$660 & $-$460$\pm$290  \\
2.7 & $-$23$\pm$6 & $-$98$\pm$6   & 7$\pm$1     & 72$\pm$7  & 62$\pm$7  & 39$\pm$1 & 870$\pm$290  & $-$1150$\pm$520 & $-$1360$\pm$300 \\
2.8 & $-$29$\pm$6 & $-$95$\pm$7   & $-$2$\pm$1  & 76$\pm$7  & 62$\pm$7  & 41$\pm$1 & 880$\pm$320  & $-$1740$\pm$580 & $-$1310$\pm$300 \\
2.9 & $-$17$\pm$6 & $-$115$\pm$7  & 5$\pm$1     & 83$\pm$7  & 68$\pm$7  & 40$\pm$1 & 840$\pm$340  & $-$820$\pm$590  & $-$1230$\pm$300 \\
3.0 & $-$7$\pm$6  & $-$126$\pm$7  & 8$\pm$1     & 90$\pm$7  & 67$\pm$8  & 42$\pm$1 & 1440$\pm$410 & $-$50$\pm$610   & $-$460$\pm$280  \\
3.1 & $-$6$\pm$7  & $-$129$\pm$8  & 1$\pm$1     & 101$\pm$8 & 67$\pm$8  & 43$\pm$1 & 1610$\pm$450 & $-$1150$\pm$690 & $-$650$\pm$290  \\
3.2 & 2$\pm$6     & $-$131$\pm$8  & $-$1$\pm$1  & 99$\pm$8  & 63$\pm$8  & 42$\pm$1 & 1470$\pm$440 & $-$1820$\pm$650 & $-$440$\pm$280  \\
3.3 & 8$\pm$7     & $-$140$\pm$8  & 0$\pm$1     & 101$\pm$8 & 66$\pm$8  & 44$\pm$2 & 2180$\pm$490 & $-$2610$\pm$740 & $-$310$\pm$300  \\
3.4 & 12$\pm$7    & $-$140$\pm$8  & 1$\pm$1     & 98$\pm$8  & 63$\pm$9  & 43$\pm$2 & 2100$\pm$480 & $-$1880$\pm$660 & 260$\pm$280     \\
3.5 & 18$\pm$7    & $-$135$\pm$9  & 5$\pm$1     & 95$\pm$9  & 63$\pm$9  & 44$\pm$2 & 1950$\pm$460 & $-$3390$\pm$740 & 260$\pm$280     \\
3.6 & 29$\pm$7    & $-$137$\pm$8  & 9$\pm$1     & 101$\pm$9 & 64$\pm$9  & 44$\pm$2 & 2290$\pm$510 & $-$2170$\pm$700 & 360$\pm$290     \\
3.7 & 7$\pm$7     & $-$132$\pm$9  & 0$\pm$2     & 91$\pm$9  & 61$\pm$10 & 44$\pm$2 & 2180$\pm$460 & $-$2150$\pm$610 & 610$\pm$280     \\
3.8 & 15$\pm$7    & $-$134$\pm$10 & 1$\pm$2     & 92$\pm$9  & 68$\pm$10 & 43$\pm$2 & 1990$\pm$450 & $-$2120$\pm$700 & $-$30$\pm$300   \\
3.9 & 5$\pm$8     & $-$139$\pm$9  & $-$6$\pm$2  & 94$\pm$10 & 66$\pm$10 & 43$\pm$2 & 1540$\pm$430 & $-$520$\pm$620  & 210$\pm$280     \\
4.0 & 7$\pm$8     & $-$143$\pm$10 & $-$1$\pm$2  & 93$\pm$9  & 66$\pm$11 & 41$\pm$2 & 1910$\pm$420 & $-$100$\pm$610  & 10$\pm$280      \\
4.1 & 17$\pm$8    & $-$145$\pm$10 & 1$\pm$2     & 92$\pm$10 & 68$\pm$10 & 44$\pm$2 & 1570$\pm$440 & 720$\pm$640     & $-$30$\pm$300   \\
4.2 & 22$\pm$8    & $-$142$\pm$11 & 2$\pm$2     & 95$\pm$10 & 72$\pm$11 & 45$\pm$2 & 1480$\pm$460 & 880$\pm$710     & 10$\pm$320      \\
4.3 & 12$\pm$9    & $-$148$\pm$10 & $-$2$\pm$2  & 94$\pm$10 & 69$\pm$11 & 48$\pm$2 & 2320$\pm$510 & 1170$\pm$680    & 180$\pm$330     \\
4.4 & 17$\pm$9    & $-$152$\pm$11 & 0$\pm$2     & 96$\pm$10 & 72$\pm$11 & 46$\pm$2 & 2330$\pm$500 & 1370$\pm$710    & $-$50$\pm$330   \\
4.5 & 19$\pm$10   & $-$158$\pm$11 & $-$2$\pm$2  & 93$\pm$11 & 76$\pm$11 & 46$\pm$2 & 2180$\pm$480 & 3960$\pm$930    & 240$\pm$360     \\
\tableline
\end{tabular}
\end{table*}


\section{RESULTS}
\label{s_res}

\subsection{Mean velocities}
\label{s_resmean}

\begin{figure}
\epsscale{1.}
\plotone{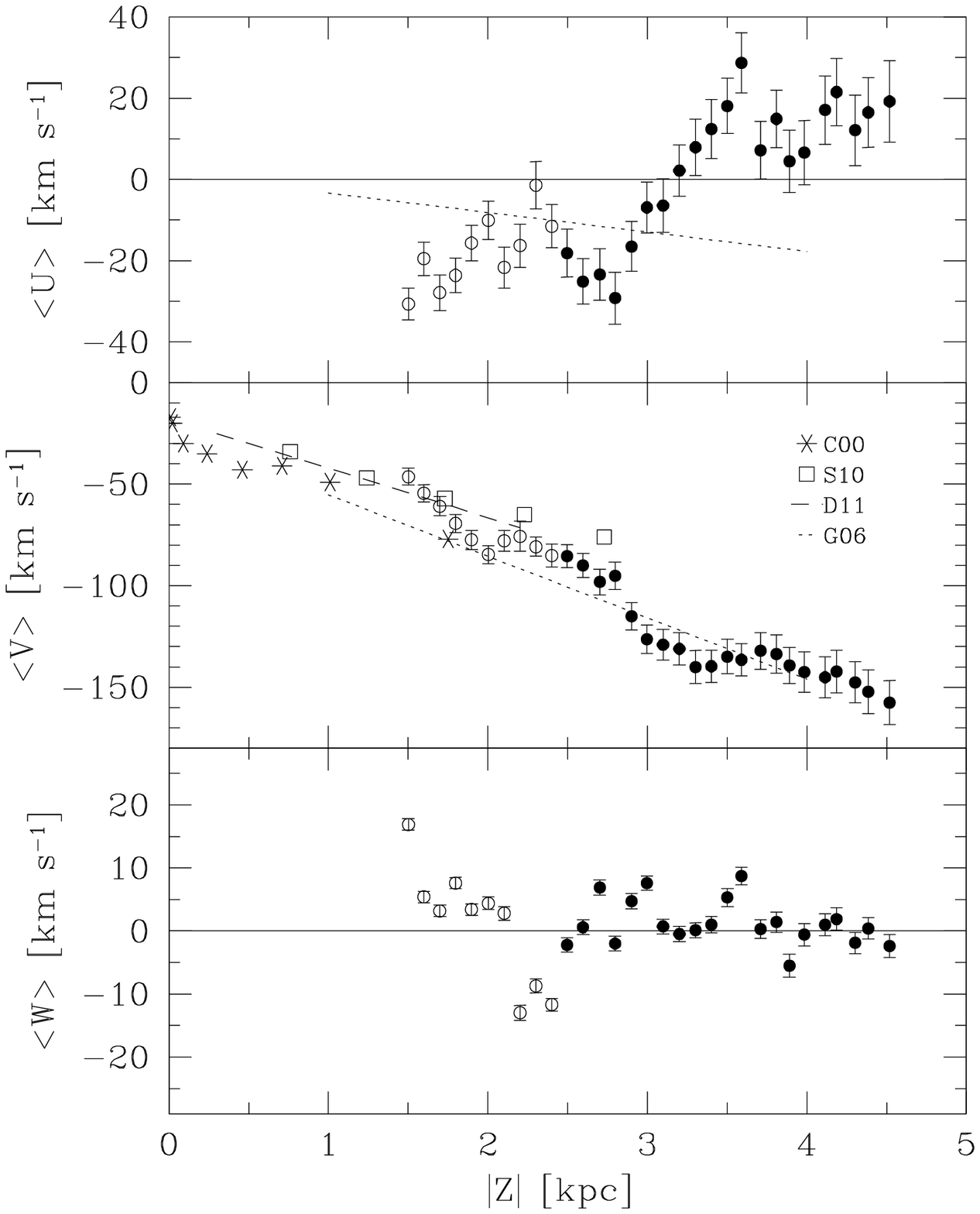}
\caption{Trend of mean velocity components (from top to bottom: radial, rotational, and vertical component) as
a function of distance from the Galactic plane. Empty dots are used for the bins contaminated by the thin disk,
where the measurements are less reliable. Results of previous investigations are also indicated:
\citet[][G06]{Girard06}, \citet[][D11]{Dinescu11}, \citet[][C00]{Chiba00}, \citet[][S10]{Spagna10}.\label{f_resmean}}
\end{figure}

The trend of the mean velocities with distance from the plane is shown in Figure~\ref{f_resmean}.
$\overline{W}$ is consistent with no significant departure from zero in the whole range. Some deviations are
observed for $\vert Z\vert\leq$2.5~kpc, but in this range the results are affected by large errors and, as already
commented, not very trustworthy. This result agrees with previous studies, that found no net vertical motion
up to various kpc from the Galactic plane \citep{Bochanski07,Bond10,Dinescu11}. On the contrary,
$\overline{U}(Z)$ has a more complex behavior: between $\vert Z \vert$=1.5 and 3~kpc we detect a non-null
mean velocity directed toward the Galactic anticenter, with an average value $\overline{U}$=$-19\pm$3~km~s$^{-1}$.
Beyond 3~kpc from the plane $\overline{U}$ abruptly increases and changes sign, and the net velocity between 3.5
and 4.5~kpc is 15$\pm$2~km~s$^{-1}$ toward the Galactic center. Another possible interpretation of the results
is that $\overline{U}$ linearly increases with $\vert Z \vert$, and a S-shaped feature is superimposed to this trend
between 2.5 and 3.5~kpc, as observed for $\sigma_\mathrm{U}$ and other kinematical quantities (see
Section~\ref{s_comove}). After the exclusion of this range, the fit returns a vertical increase of
15$\pm 2$~km~s$^{-1}$~kpc$^{-1}$.
Adopting the new values of \citet{Schonrich10} for the solar peculiar motion causes not a relevant change on the
results, as $\overline{U}$ would be higher by about 1~km~s$^{-1}$ only.

Previous studies have shown evidence that stars more distant than 1~kpc from the plane show a non-null net
radial motion of the order of $\sim$9~km~s$^{-1}$ toward the Galactic anticenter
\citep{Smith09,Rybka10,Dinescu11}. Our results between $\vert Z \vert$=1.5 and 3~kpc agree with their conclusion,
although our mean value is larger by about a factor of two. The sign flip observed at larger $\vert Z \vert$, on the
contrary, has never been reported in the literature. Nevertheless, \citet{Smith09} analyze
halo stars, \citet{Rybka10} do not reach these heights above the plane, and \citet{Dinescu11} have too
few stars in this range. Very interestingly, \citet{Bond10} detect a small positive mean radial component out
to $\vert Z\vert\approx 3$~kpc, and a negative value in their more distant bins, although the authors conclude that
these results are consistent, within errors, with a null net motion. \citet{Dinescu11} proposed that an
inward motion of the Local Standard of Rest (LSR) of the order of 10~km~s$^{-1}$ is responsible for the
observed non-null mean value of $\overline{U}$, as suggested by the fact that this is observed among both
disk and halo stars more distant than 1~kpc from the sun. This hypothesis cannot account for any vertical
trend of $\overline{U}$ other than a constant non-zero value at any $\vert Z \vert$. Thus, while it is not contradicted
by our results, at least another effect must be invoked. For example, an outward motion of
$\sim$25~km~s$^{-1}$ of the stars between 3.5 and 4.5~kpc, coupled with the mentioned LSR motion, could
explain our results. This clumpy kinematical distribution would not be surprising, because perturbations
produced by the bar and the spiral arms are expected to form groups of stars with velocity offset as large
as 50~km~s$^{-1}$ \citep{Quillen10}.

\citet{Girard06} studied the proper motions of our same stellar sample, and their results for
$\overline{U}(Z)$ are overplotted to ours in the upper panel of Figure~\ref{f_resmean}, after correcting for
the solar peculiar motion and changing the sign of $U$ to account for the different direction of the reference
axis. The agreement with our results is poor: while their nearly flat profile roughly coincides with our mean
value ($-$7.7~km~s$^{-1}$), they did not detect any change of sign, nor a steep positive gradient. The
different approach to the same data must have caused the different results. For example the features observed
by us could have been masked in \citet{Girard06} by their smoothed, statistical distance determination, or by
dwarf stars and halo contaminants, removed in our work.

\begin{figure}
\epsscale{1.}
\plotone{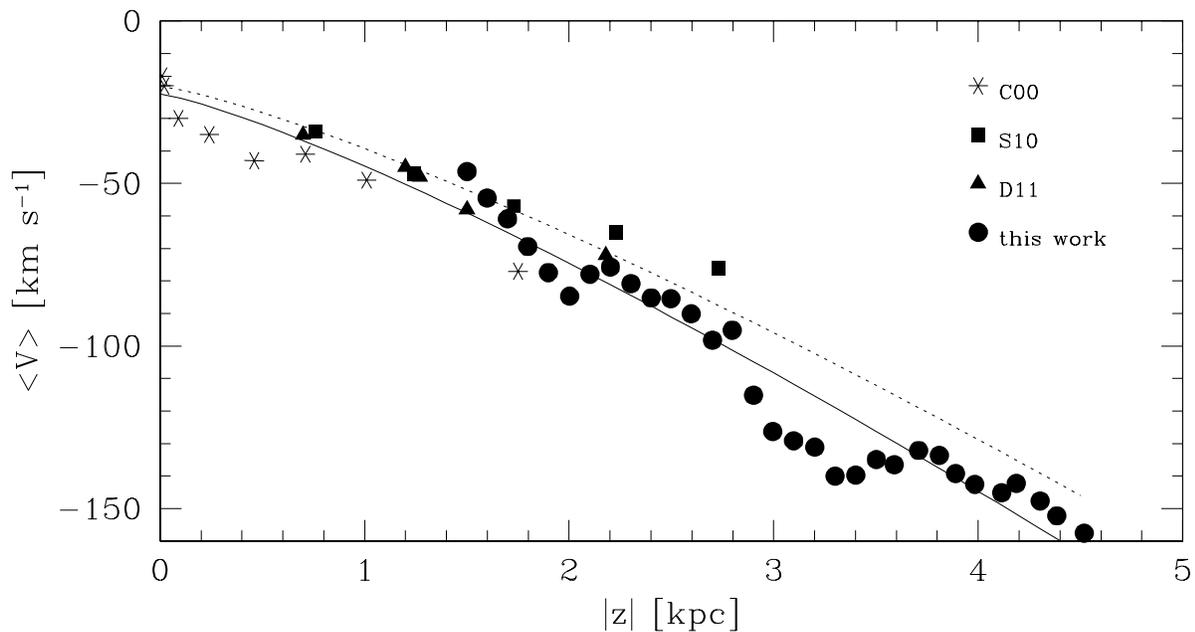}
\caption{Rotational velocity of thick disk stars as a function of distance from the plane. Full dots: our
work; asterisks: \citet[][C00]{Chiba00}; squares: \citet[][S10]{Spagna10}; triangles: \citet[][D11]{Dinescu11}.
The dotted curve indicates the power-law solution proposed by \citet{Bond10}, the thick curve is the analogous
relation obtained from the fit of the plotted data points.\label{f_Vall}}
\end{figure}

In the middle panel of Figure~\ref{f_resmean} we show the measured trend of $\overline{V}(Z)$, compared
to other results from the literature. The mean rotational velocity of thick disk stars decreases with
distance from the plane, due to the well-known vertical shear first detected by \citet{Majewski92}.
The fit of the data, after the exclusion of the less reliable bins at $\vert Z\vert\leq$2.5~kpc, yields
$\overline{V}(Z)=(-25\pm 12)-(30\pm 3)\cdot \vert Z\vert$~(km~s$^{-1}$). As can be seen in
Figure~\ref{f_resmean}, this
solution excellently matches the results of \citet{Girard06}, and even the data points of \citet{Chiba00}, at
$z\leq 2$~kpc, are well aligned with the derived linear relation. Both these investigations measure a
vertical shear of $-$30~km~s$^{-1}$~kpc$^{-1}$, as also recently found by \citet{Brown08}, and
\citet{Bond10}. Had we fitted all the data points down to $\vert Z \vert$=1.5~kpc, we would have found a steeper
slope ($-35.1\pm 1.8$~km~s$^{-1}$~kpc$^{-1}$), at the upper limit of the range spanned by literature values,
which vary from $-16\pm$4 \citep{AllendePrieto06} to $-36\pm$1~km~s$^{-1}$ \citep{Carollo10}.
The revised values for the solar peculiar motion proposed by \citet{Schonrich10} cause an upward revision
of the results by 7.1~km~s$^{-1}$.

\citet{Dinescu11} and \citet{Spagna10} measured a shallower slope ($-25.2\pm$2.1 and
$-19\pm$2~km~s$^{-1}$~kpc$^{-1}$, respectively) between 0.7 and 2.8~kpc. While our results are compatible
with theirs in our nearest bins, we find a lower mean rotational velocity beyond $\vert Z \vert$=2~kpc. Nevertheless,
the different vertical rotational gradient found by these studies is not necessarily a disagreement,
because they sample a different $\vert Z \vert$-range, and the underlying shear is not
required to be strictly linear. In fact, \citet{Ivezic08} have proposed the non-linear relation
$\overline{V}(Z)=-20.1-19.2\cdot \vert Z\vert^{1.25}$~km~s$^{-1}$ from the analysis of SDSS data. The
combined data points of \citet{Chiba00}, \citet{Spagna10}, \citet{Dinescu11}, and of the present work, closely
follow this equation (see Figure~\ref{f_Vall}), and a fit in the form
$\overline{V}(Z)=\alpha+\beta\cdot \vert Z\vert^{\gamma}$~km~s$^{-1}$ returns a very similar solution,
with $\alpha =-22.5$~km~s$^{-1}$, $\beta =-22.2$~km~s$^{-1}$~kpc$^{-1}$, and $\gamma =$1.23. Very noticeably,
the results of four surveys finding a different linear relation are all well reproduced by a single
non-linear curve proposed by an independent work. The underlying vertical trend of the thick disk
rotational velocity is therefore most probably close to but not exactly linear, and better represented by
a low-order power law.

\subsection{Velocity dispersions}
\label{s_resdisp}

\begin{figure}
\epsscale{1.}
\plotone{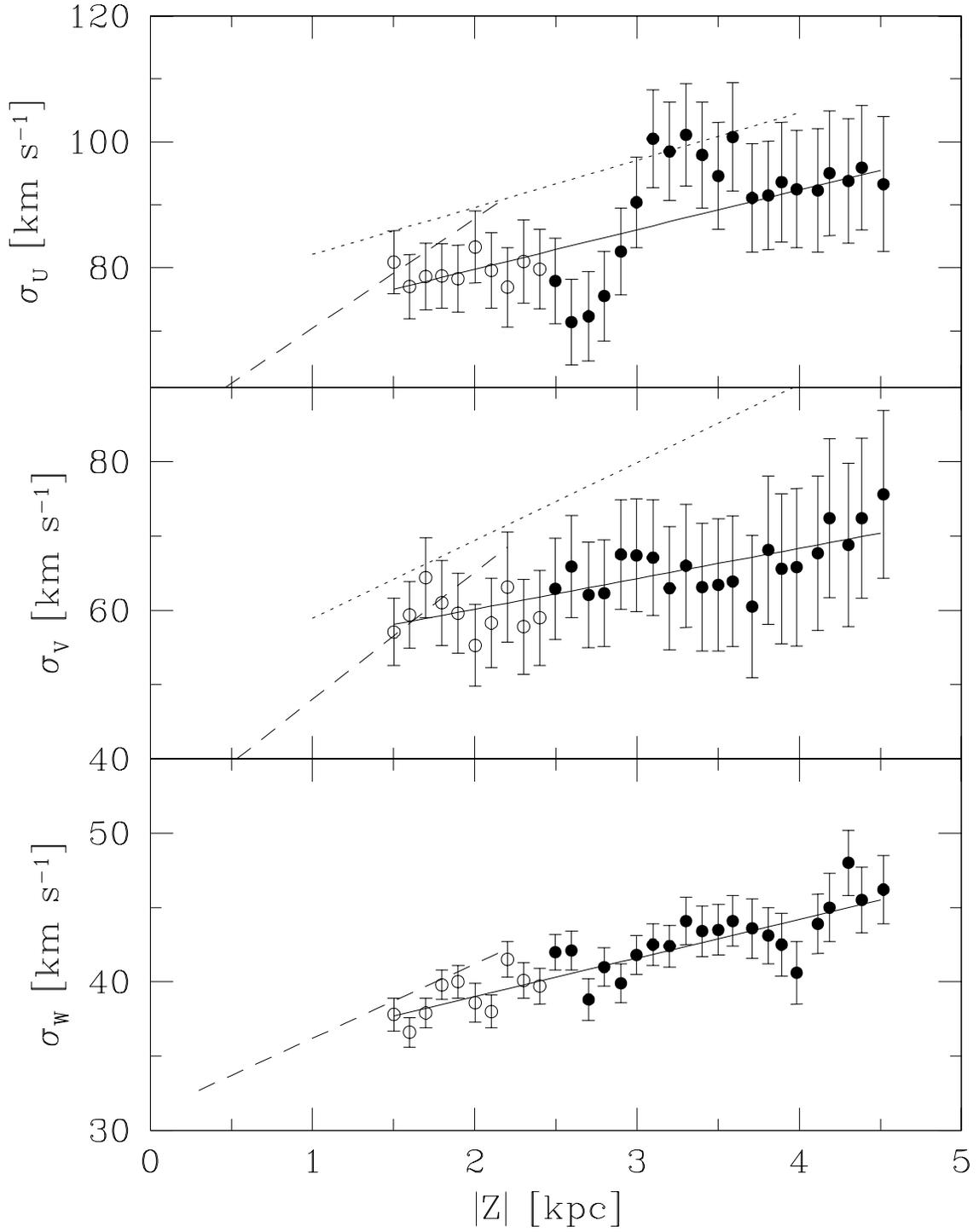}
\caption{Velocity dispersion as a function of distance from the plane (from top to bottom: radial,
rotational, and vertical velocity dispersion). The thick curve indicates the linear fit of the data given
in Equations~(\ref{e_sU}) to (\ref{e_sW}). The empty dots, dashed and dotted lines are as
Figure~\ref{f_resmean}.\label{f_disp}}
\end{figure}

The vertical profile of the velocity dispersions is shown in Figure~\ref{f_disp}, compared to other
works from the literature. We excluded from comparison the results of \citet{Bond10}, because they do
not distinguish between different disk components, thus finding steeper gradients as a result of the
mix of thin and thick disk stars. Caution must also be taken when comparing the results obtained in
different ranges of $\vert Z \vert$, because the underlying trend of the dispersions is not necessarily linear, and
the gradient can assume different values. For example, the models of \citet{Girard06} indicate
that the vertical profile should be progressively flatter at increasing $\vert Z \vert$.

The linear fit of the trends with $\vert Z \vert$ yields:
\begin{equation}
\sigma_\mathrm{U}(Z)=(82.9\pm 3.2)+(6.3\pm 1.1)\cdot(\vert Z \vert-2.5) \mathrm{km~s^{-1}},
\label{e_sU}
\end{equation}
\begin{equation}
\sigma_\mathrm{V}(Z)=(62.2\pm 3.1)+(4.1\pm 1.0)\cdot(\vert Z \vert-2.5) \mathrm{km~s^{-1}},
\label{e_sV}
\end{equation}
\begin{equation}\sigma_\mathrm{W}(Z)=(40.6\pm 0.8)+(2.7\pm 0.3)\cdot(\vert Z \vert-2.5) \mathrm{km~s^{-1}},
\label{e_sW}
\end{equation}
where $\vert Z \vert$ is in kpc. The quoted errors were obtained restricting the fit to a subset of eight
non-overlapping bins, to avoid the underestimate induced by the correlation between the data points.
The results are identical to those of \citet{Moni10}, despite the different binning scheme, except for
$\sigma_\mathrm{V}$ for which we derive a slightly smaller gradient, but the difference is not
significant (0.4~km~s$^{-1}$~kpc$^{-1}$). The gradients are small, and the three dispersions increase
by only $\sim$7\% between 2.5 and 3.5~kpc. This explains why the change of the thick disk kinematics
with distance from the plane has not been clearly detected for about two decades after its discovery.

The results for $\sigma_\mathrm{W}$ are the most precise and reliable, because $\sim$90\% of the
information on $W$ comes from RVs, whose errors are an order of magnitude smaller than those in proper
motions. $\sigma_\mathrm{W}(Z)$ shows a small but clear and constant increase, with small scatter
around the derived linear relation. The vertical gradient is smaller than the one found by
\citet{Dinescu11} by about a factor of two, but the results are consistent at the 1$\sigma$ level.
\citet{Yoss87} propose an even steeper profile ($\sim$10~km~s$^{-1}$~kpc$^{-1}$) in the range
$\vert Z \vert$=0--2~kpc, but the same authors suspect that this could be due to contamination by halo stars
increasing with Galactic height.

The dispersion of the rotational velocity component, $\sigma_\mathrm{V}$, also shows a clear increase
with $\vert Z \vert$, although the data points are affected by larger errors and are more scattered. The vertical
gradient is smaller than the results of both \citet{Dinescu11} and \citet{Girard06}, by a factor of
three and two, respectively, and the difference is at the 2$\sigma$ level in both cases.

The vertical profile of $\sigma_\mathrm{U}$ shows a very peculiar behavior, with large deviations from
linearity between 2.5 and 3.5~kpc. This feature will be discussed in Section~\ref{s_comove}. Outside
this range, the data points closely follow a linear relation, and the solution given in
Equation~\ref{e_sU} was obtained excluding this interval from the fit. The derived
vertical gradient of $\sigma_\mathrm{U}$ is very similar to that found by \citet{Girard06}, but their
solution is offset toward higher values by about 10~km~s$^{-1}$, while the gradient measured by
\citet{Dinescu11} is 4$\sigma$ times higher than ours.

In conclusion, we confirm that the velocity dispersions of the Galactic thick disk increase with
distance from the plane, as suggested by previous investigations
\citep{Yoss87,Yoss97,Soubiran03,Girard06,Ivezic08,Dinescu11}, but we derive vertical gradients that
are in general smaller than other studies. The differences with \citet{Dinescu11} can be at least in
part due to the aforementioned expected decrease of the gradient with distance from the plane, because
the results for $\sigma_\mathrm{W}$ are consistent, and the solutions proposed for the other two
components overlap in the $Z$-range in common (1.5--2.2~kpc). On the contrary, \citet{Girard06}
studied our same sample, and the different data analysis must be the cause of the disagreement. Since
they measured both higher dispersions and steeper vertical gradients than us, their results could have
been affected by the dwarf stars and halo contaminators removed by us, and/or the thin disk
contamination, that they did not take into account.

\begin{figure}
\epsscale{1.}
\plotone{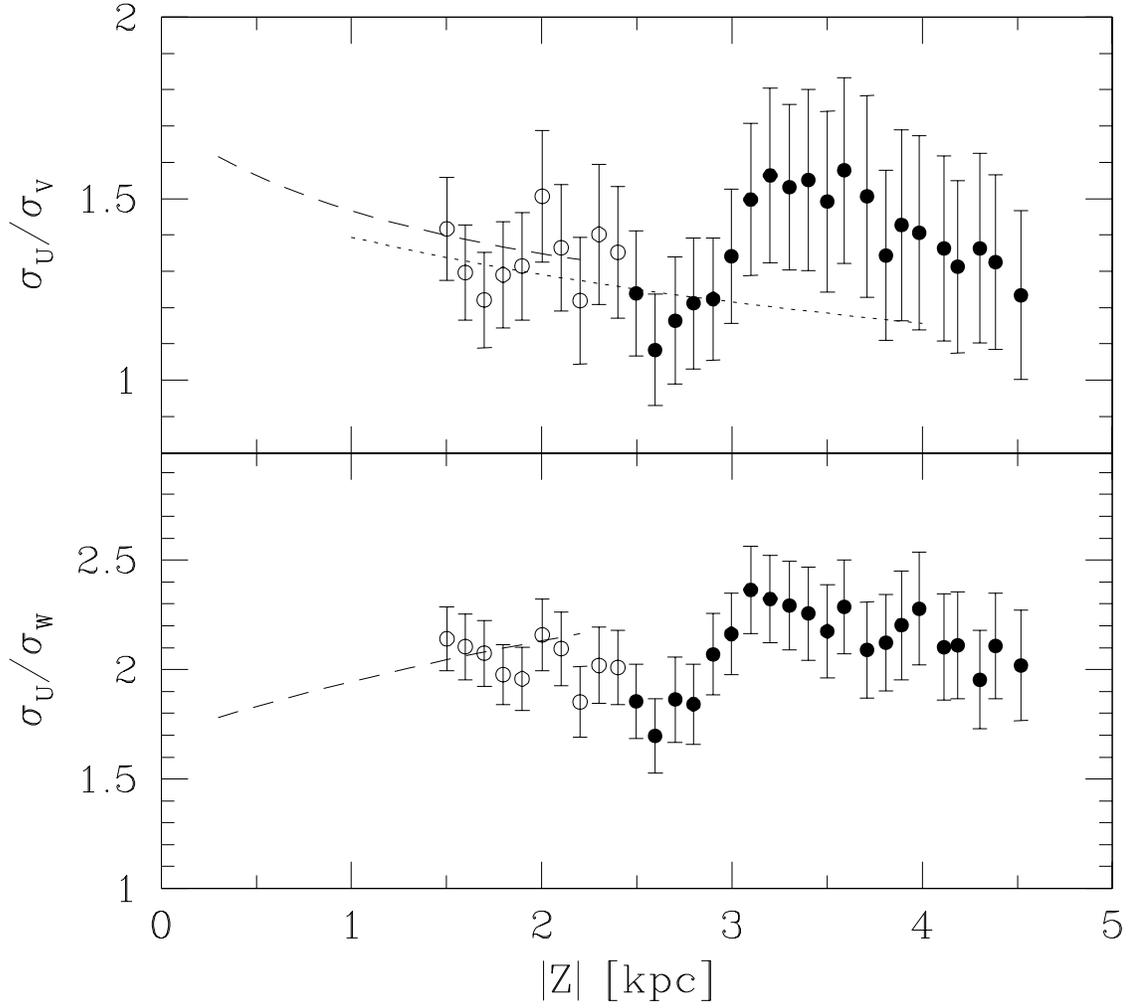}
\caption{Ratio of dispersions as a function of distance from the plane. Upper panel:
$\sigma_\mathrm{U}/\sigma_\mathrm{V}$; lower panel: $\sigma_\mathrm{U}/\sigma_\mathrm{W}$. The empty
dots, dashed and dotted curves are as in Figure~\ref{f_resmean}.\label{f_ratio}}
\end{figure}

The three dispersions increase with distance from the plane proportionally, and the anisotropy is
approximatively constant at all $\vert Z \vert$. In Figure~\ref{f_ratio} the vertical profile of the ratios
$\sigma_\mathrm{U}$/$\sigma_\mathrm{W}$ and $\sigma_\mathrm{U}$/$\sigma_\mathrm{V}$ is shown, where
the errors come from propagation of the uncertainties on the dispersions.

The mean value of $\sigma_\mathrm{U}$/$\sigma_\mathrm{W}$ is 2.08$\pm$0.06, where the error was
estimated from the statistical error on the mean for eight uncorrelated bins, as before. The linear fit
of the data points returns a negligible gradient (0.09$\pm$0.07~kpc$^{-1}$), and the data are consistent
with a flat profile. Literature values for $\sigma_\mathrm{U}$/$\sigma_\mathrm{W}$ span a wide range
from 1.1 \citep{Veltz08} to 1.9 \citep[e.g.,][]{Vallenari06}, and our results are at the upper end of
this distribution. As shown in the upper panel of Figure~\ref{f_ratio}, the results of \citet{Dinescu11}
agree with our measurements in the Z-range in common, but they deviate from our measurements if
extrapolated to higher distance from the plane.

The results for $\sigma_\mathrm{U}$/$\sigma_\mathrm{V}$ also show no significant gradient
(0.06$\pm$0.05~kpc$^{-1}$) and a mean value of 1.36$\pm$0.05. This is lower than the value predicted by
the epicyclic approximation \citep[1.49, cfr. Equation 3-76 of][]{Binney98}, indicating that the population
under analysis cannot be assumed in nearly circular orbits. Our results are well within the range spanned
by literature values, which vary from $\approx$1 \citep[e.g.,][]{Chiba00,Carollo10}, to $\approx$1.6
\citep{Soubiran03}. As in the case of $\sigma_\mathrm{U}$/$\sigma_\mathrm{W}$, the results of
\citet{Dinescu11} overlap with our data points in the range in common, but their extrapolation to higher
$\vert Z \vert$ do not. The analytical expressions given by \citet{Girard06} for $\sigma_\mathrm{U}$ and
$\sigma_\mathrm{V}$, on the contrary, return a value of their ratio much lower than our data points at
any Galactic height. This is probably due to their much higher $\sigma_\mathrm{V}$, as shown in the middle
panel of Figure~\ref{f_disp}.

\subsection{Orientation of the dispersion ellipsoid}
\label{s_resangles}

\begin{figure}
\epsscale{1.}
\plotone{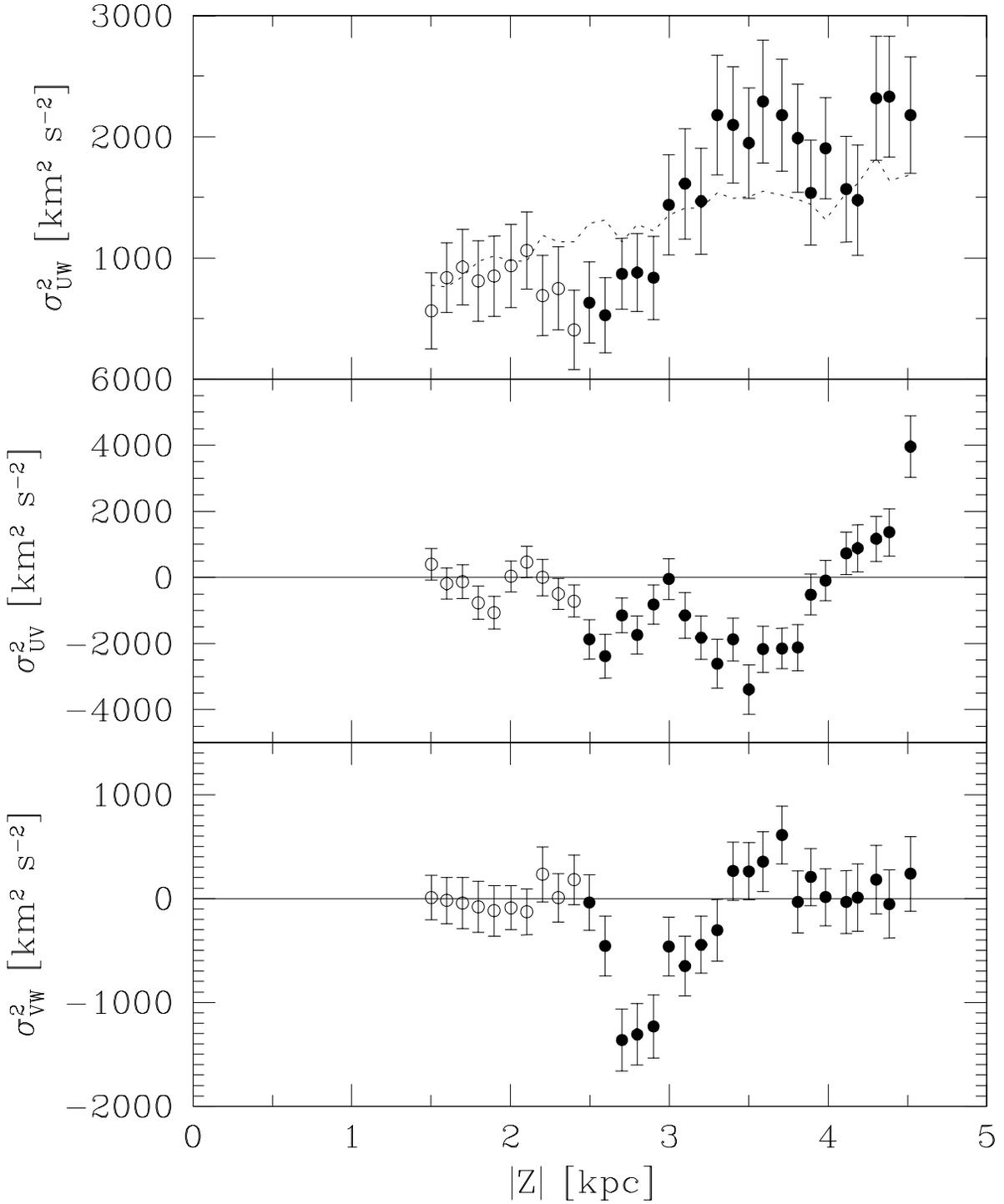}
\caption{Vertical trend of the non-diagonal terms of the dispersion matrix, $\sigma^2_\mathrm{UW}$,
$\sigma^2_\mathrm{UV}$, and $\sigma^2_\mathrm{VW}$ (from top to bottom). The empty dots are used for bins
contaminated by the thin disk. In the upper panel, the dashed line indicates the trend of the analytical
expression proposed by \citet{Kuijken89}.
\label{f_cross}}
\end{figure}

The non-diagonal term of the dispersion matrix $\sigma^2_\mathrm{ij}$ gives a measurement of the
correlation between the $i$-th and $j$-th velocity components. In fact, if two principal axis of the
dispersion ellipsoid are aligned with the $i$-th and $j$-th axis, we have $\sigma^2_\mathrm{ij}$=0, and
the two velocity components are uncorrelated. The orientation of the dispersion ellipsoid of an old,
dynamically relaxed population is related to the shape of the Galactic potential
\citep{LyndenBell62,Ollongren62,Hori63}, and can indicate the presence of non-axisymmetric structures in
the disk \citep{Kuijken91}. Moreover, \citet{Bienayme09} and \citet{Siebert08} have shown that the
vertical tilt of the ellipsoid allows to estimate the flattening of the dark halo.
However, calculating the expectation of a specific halo model through integration of orbits is beyond
the scope of the paper, and here we will only present the observational results, for use of future works.

\begin{figure}
\epsscale{1.}
\plotone{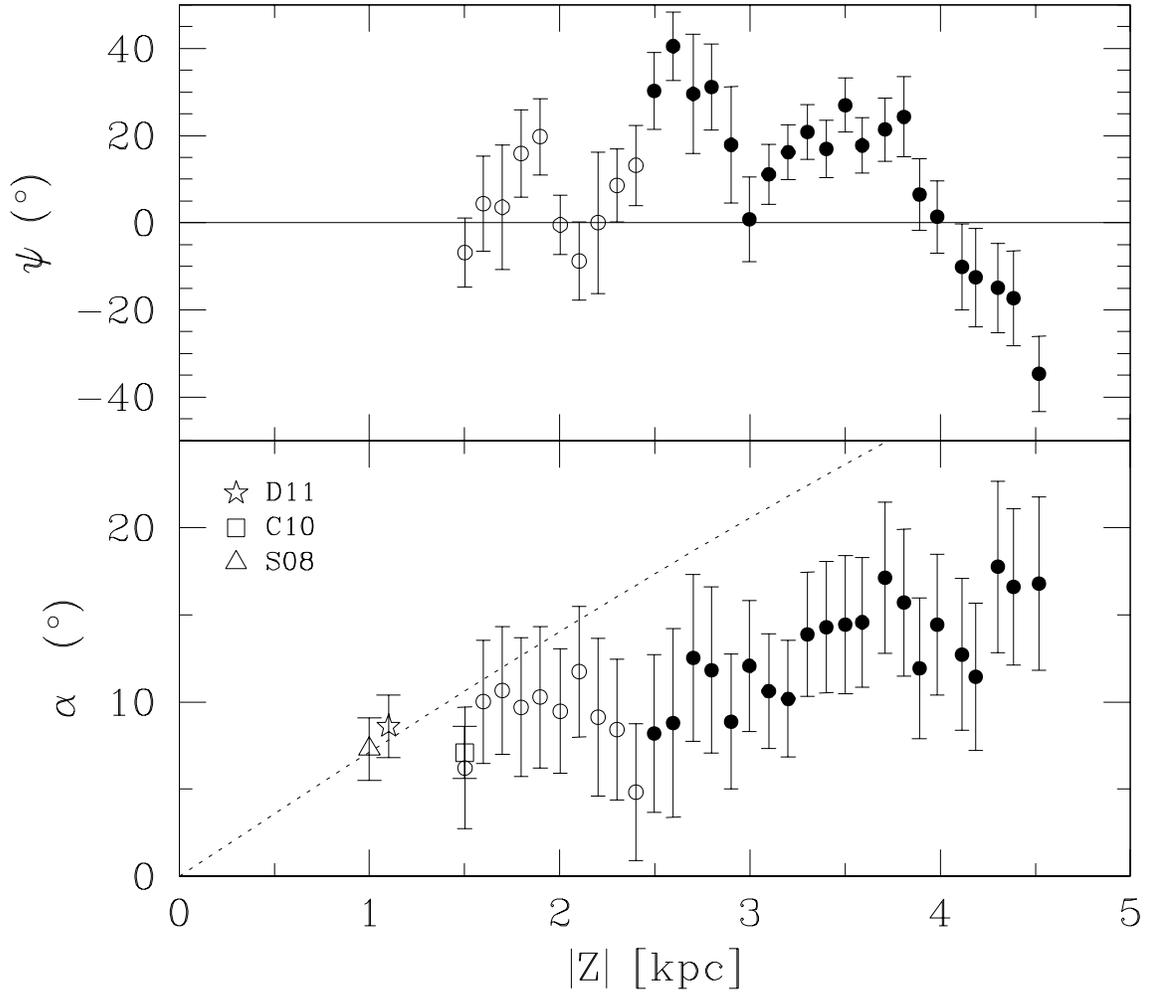}
\caption{Measured vertex deviation (upper panel) and tilt angle (lower panel) as a function of distance
from the plane. The empty dots are used for the bins contaminated by thin disk. In the lower panel the
dotted line shows the tilt angle of a dispersion ellipsoid aligned with the spherical coordinate system at
any $\vert Z \vert$, and previous literature measurements are also shown:
\citet[][D11]{Dinescu11}, \citet[][C10]{Carollo10}, and \citet[][S08]{Siebert08}.\label{f_angles}}
\end{figure}

Our results for the three cross-terms are shown in Figure~\ref{f_cross}. The profile of
$\sigma^2_\mathrm{UV}$ closely follows a decreasing linear relation up to $\vert Z \vert$=4~kpc, but abruptly jumps
to positive values at larger heights. This is probably due to the sign flip of $\overline{U}$ analyzed
in Section~\ref{s_comove}. In any case, the results indicate that the $U$ and $V$ velocity components
are correlated, and the velocity ellipsoid is titled in the radial-longitudinal plane. The rotation angle,
i.e. the vertex deviation $\psi$, was calculated in each bin by means of the relation:
\begin{equation}
\psi=-\frac{1}{2}arctg\Bigl{(}\frac{2\sigma^2_\mathrm{UV}}{\sigma^2_\mathrm{U}-\sigma^2_\mathrm{V}}\Bigl{)}
\label{e_vertex}
\end{equation}
\citep{Amendt91}. The results are shown in the upper panel of Figure~\ref{f_angles}. We measure a
non-negligible vertex deviation, increasing from nearly zero at $\vert Z \vert$=1.5~kpc to $\sim 20\degr$ at 3.5~kpc.
The linear fit in the range 1.5--4~kpc yields the relation $\psi =-1.0+5.4\cdot \vert Z \vert$. Our results agree with
\citet{Dinescu11}, who measured $\psi=8.2\pm3.2\degr$ at $Z$=1.1~kpc. Previous investigations evidenced
that the vertex deviation decreases from $\approx 20\degr$ for young populations to nearly zero for old,
metal-poor disk stars \citep{Bienayme99,Dehnen98,Soubiran03,Fuchs09}. Our results are not at variance
with this conclusion if the vertex deviation increases with $\vert Z \vert$, as suggested by our observations,
because previous studies were limited to small Galactic heights.

$\sigma^2_\mathrm{VW}$ shows no significant deviation from zero in the whole range of $\vert Z \vert$, except between
2.5 and 3.5~kpc, as discussed in Section~\ref{s_comove}. This indicates that the $V$ and $W$ velocity
components are not correlated. On the contrary, $\sigma^2_\mathrm{UW}$ is significantly different from
zero, and steadily increases with $\vert Z \vert$. Proposing a reliable linear expression for $\sigma^2_\mathrm{UW}(Z)$
is not straightforward, because it shows irregular fluctuations. As expected by \citet{Binney83} and
\citet{Binney98}, $\sigma^2_\mathrm{UW}$ is always bound between zero and
$\sigma^2_\mathrm{UW,max}=(\overline{U^2}-\overline{W^2})\cdot (Z/R)$, the value
assumed when the velocity ellipsoid is aligned with the spherical coordinate system, because we find that
$\sigma^2_\mathrm{UW,max}\geq 4500$~km$^{2}$~s$^{-2}$ at any Galactic height. We also find that the
expression proposed by \citet{Kuijken89}, obtained under the assumption that the ellipsoid points toward
the Galactic center and its axis ratio is constant in the spherical coordinate system, is a relatively
good approximation of the measured value up to about 3~kpc from the plane, but the predicted slope is too
shallow and the agreement with observations degrades with $\vert Z \vert$ (see Figure~\ref{f_cross}).

The tilt angle in the $U$-$W$ plane can be calculated, analogously to the vertex deviation, from the
Equation:
\begin{equation}
\alpha=-\frac{1}{2}arctg\Bigl{(}\frac{2\sigma^2_\mathrm{UW}}{\sigma^2_\mathrm{U}-\sigma^2_\mathrm{W}}\Bigl{)},
\label{e_alpha}
\end{equation}
and the results are shown in Figure~\ref{f_angles}. We measure a slight increase of $\alpha$ with
$\vert Z \vert$, and the linear fit of the data, after the exclusion of two deviating points, yields the relation
$\alpha (\vert Z \vert)=9\fdg 6+2\fdg 4\cdot [(\vert Z \vert/kpc)-2)]$. The results agree with previous works at
$\vert Z \vert$=1--1.5~kpc \citep{Siebert08,Carollo10,Dinescu11}, but not with the measurements of \citet{Fuchs09}
at $\vert Z \vert\leq$1~kpc, whose sample is most probably dominated by thin disk stars. The dashed curve in
Figure~\ref{f_angles} indicates the value of $\alpha$ when the dispersion ellipsoid points toward the Galactic center,
assuming R$_\odot$=8~kpc. The tilt angle is constantly lower than this in the range 1.5--4.5~kpc, and the ellipsoid
thus is directed toward a point located behind the Galactic center, at a Galactocentric distance increasing
with $\vert Z \vert$ from R$_o$=2.3 to 9~kpc. Orbit integration studies indeed predict this result
\citep{Kuijken89,Binney83,Kent91,Shapiro03}, but the inclination is noticeably higher than the
expectations, because theoretical calculations return R$_o$=5--10~kpc at $\vert Z \vert$=1.1~kpc. This could indicate
that the Galactic potential used in these studies needs to be refined to match the observations. On the
contrary, the increase of $\alpha (\vert Z \vert)$ modeled by \citet{Bond10} is too steep, and their expectation
$\alpha$(3.5~kpc)=26$\degr$ is at variance with our results.

\subsection{MOdified Newtonian Dynamics}
\label{s_mond}

\citet{Bienayme09b} showed that the vertical trend of the tilt angle is an excellent observational
signature of the underlying gravity law, and it can be used as a test for the MOdified Newtonian Dynamics
theory \citep[MOND,][]{Milgrom83}, because its expectation diverges from that of the Newtonian dynamics with
distance from the plane. Unfortunately, the calculations of \citet{Bienayme09b} are limited to $\vert Z \vert\leq$2~kpc,
and the range of overlap with our data is very narrow. The value of $\alpha$ expected by \citet{Bienayme09b} at
$\vert Z \vert$=2~kpc is 12$\degr$ for Newtonian dynamics and 10$\degr$ for MOND. We find
$\alpha$(2~kpc)=$9\fdg 5\pm 3\fdg 6$ and the mean of the five measurements in the range 1.8-2.2~kpc is
$10\fdg 0\pm0.5\degr$, in excellent agreement with MOND expectations. The linear fit presented in
Section~\ref{s_resangles} also implies $\alpha$(2~kpc)=$9\fdg 6$. However, we derive a vertical gradient
($2\fdg 4$~kpc$^{-1}$) much shallower than the predictions of \citet{Bienayme09b} for both MOND
(5$\degr$~kpc$^{-1}$) and Newtonian dynamics (6$\degr$~kpc$^{-1}$). In conclusion, our observations agree better with
the models that \citet{Bienayme09b} derived for MOND than with those from Newtonian dynamics, but the extension of
\citet{Bienayme09b} calculations to higher $\vert Z \vert$, and to specific initial conditions for the thick disk, are
required to perform a reliable test of the gravitational law.

\subsection{Subtructures in the thick disk}
\label{s_comove}

The vertical profile of $\sigma_\mathrm{U}(Z)$ shows a puzzling behavior between 2.5 and 3.5~kpc from the
Galactic plane (top panel, Figure~\ref{f_disp}), where the dispersion deviates from the linear trend. A
similar feature is clearly observable in the trend of other quantities related to the radial velocity component $U$,
like $\overline{U}$ and $\sigma^2_\mathrm{UW}$ \citep[as already noted by][]{Moni10}, but deviations from the linear
trend in the range $\vert Z \vert$=2.5--3~kpc could be present even in other profiles, such as $\overline{V}(Z)$,
$\overline{W}(Z)$, $\sigma_\mathrm{W}(Z)$, and $\sigma^2_\mathrm{VW}(Z)$. The origin of this behavior is unclear.
It is possible that a group of comoving stars, forming a sub-structure in the Galactic thick disk, is affecting the
measured kinematics between 2.5 and 3.5~kpc from the plane. In particular, the similarity of the profile of two
totally unrelated quantity such as $\sigma^2_\mathrm{VW}(Z)$ and $\sigma_\mathrm{U}(Z)$ is instructive, because
it excludes the possibility that this behavior is only due to some bad measurements. In any case, the existence of a
kinematical substructure among our stars cannot be claimed on the basis of these results only.

\subsection{Radial behavior of $\sigma_\mathrm{U}$}
\label{s_radial}

\begin{figure}
\epsscale{1.}
\plotone{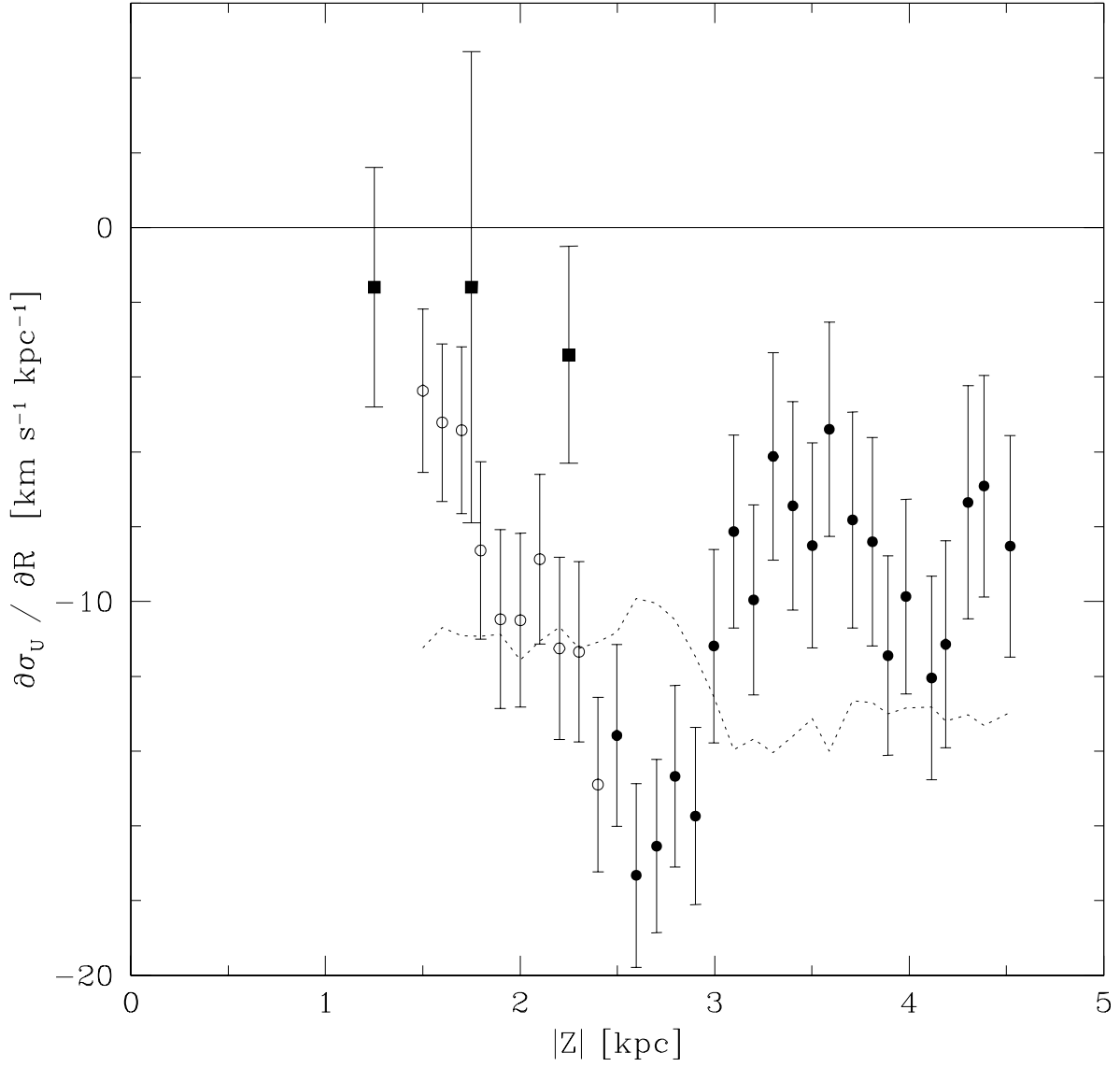}
\caption{Estimated radial derivative of $\sigma_\mathrm{U}$ as a function of distance from the plane.
The empty dots are used for the bins contaminated by thin disk. The squares correspond to the results of
\citet{Dinescu11}. The dotted line indicates the theoretical expectation assuming a radially constant anisotropy.
\label{f_jeans}}
\end{figure}

Our targets are distributed in a very narrow range of $R$, and the data do not provide any direct information
about the radial behavior of the kinematical quantities. Nevertheless, some indirect indication can be derived.
Manipulating the radial Jeans equation in cylindrical coordinates in steady state, with the radial component
of the force expressed as F$_\mathrm{R}=-v^2_c/\mathrm{R}$, we obtain:
\begin{equation}
\frac{\partial \sigma^2_\mathrm{U}}{\partial R}=\frac{\overline{UW}}{h_{Z,\rho}}-\frac{\partial \overline{UW}}{\partial Z}
-\frac{v^2_c}{R}+\frac{\sigma^2_\mathrm{V}+\overline{V}^2}{R}-\sigma^2_\mathrm{U}\Bigl{(}\frac{1}{R}-\frac{1}{h_{R,\rho}}\Bigl{)},
\label{e_jeans}
\end{equation}
where we also assumed that $\overline{\mathrm{U}}^2\ll \sigma^2_\mathrm{U}$,
$\partial \overline{\mathrm{U}}$/$\partial R$=0, and that the density decays exponentially with both $R$ and $Z$,
with scale length $h_{R,\rho}$ and scale height $h_{Z,\rho}$. Inserting the observed quantities in the right hand
side of Equation~\ref{e_jeans} we can thus estimate, as an exercise, the radial behavior of $\sigma_\mathrm{U}$.
The large uncertainties involved prevent a precise measurement, but the comparison with the expectations provides
a consistency check for the kinematical data here presented, because an unphysical result could indicate a problem
with them. The results are shown in Figure~\ref{f_jeans}, where we assumed $h_{R,\rho}$=3.6~kpc and
$h_{Z,\rho}$=0.9~kpc \citep{Juric08}, R$_\odot$=8~kpc, and $v_c$=220~km~s$^{-1}$. The data points are scattered around
the mean value $-10\pm 1$~km~s$^{-1}$~kpc$^{-1}$, following the behavior of $\sigma_\mathrm{U}(Z)$ discussed in
Section~\ref{s_comove}, with no clear vertical trend. In the nearest bins the results are partially consistent, but
more negative, than what found by \citet{Dinescu11} and \citet[][$-3.8\pm 0.6$~km~s$^{-1}$~kpc$^{-1}$]{Neese88},
although Neese \& Yoss refer to the thin disk. The observations show \citep{Kruit81,Kruit82} that $\sigma_\mathrm{W}$
should exponentially decay in the radial direction, with the same scale length of the mass density. This is often
considered valid even for $\sigma_\mathrm{U}$, under the assumption of a constant anisotropy, and \citet{Cuddeford92}
found that this should be the best approximation at the solar position. The results also roughly agree with the
expectation of this model, as shown in Figure~\ref{f_jeans}. In conclusion, the presented kinematics implies a radial
behavior of $\sigma_\mathrm{U}$ consistent with both previous observations and with the expectations of theoretical
predictions.

\subsection{Comparison with models of thick disk formation}
\label{s_models}

The presence of kinematical gradients in the Galactic thick disk is a powerful diagnostic to discriminate between the
various models of its formation \citep{Majewski93}, but unfortunately the model expectations concerning the vertical
gradient have not been investigated so far. It would be interesting to compare, in the near future, the predictions
of the various models with the observed trend with $Z$ of the velocity dispersions. On the contrary, the radial
gradient of the dispersions has been modeled in the context of thick disk formation through disk heating by the
merging of minor satellites \citep{Villalobos08,Bekki11}. The observations of \citet{Dinescu11} agree with the
results of these simulations, provided that a low inclination orbit of the merging satellite is assumed. A similar
conclusion is drawn comparing our measurements of $\sigma_\mathrm{U}$/$\sigma_\mathrm{W}$ with the theoretical
expectations of the merging scenario. In fact, the simulations showed that this ratio is strongly linked to the
inclination angle of the orbit \citep{Villalobos08,Purcell09}, and our result
($\sigma_\mathrm{U}$/$\sigma_\mathrm{W}$=2.08) favors a small inclination angle, $i\approx$0--30$\degr$
\citep[see Figure~15 of][]{Villalobos10}. A low inclination orbit of the infalling satellite is also required to
reproduce the rather large vertical gradient of the rotational velocity, found in this work as in previous
investigations \citep{Villalobos08,Bekki11}. On the contrary, the radial migration model predicts a much shallower
gradient \citep[$-$17~km~s$^{-1}$~kpc$^{-1}$,][]{Loebman11}, incompatible with the observations.

In conclusion, the models of thick disk formation have so far provided only fragmentary predictions about its
kinematical properties, and our observations cannot still be used to fully discriminate between them. However, we
find that our results are consistent with the scenario where the thick disk formed through dynamical heating of a
pre-existing Galactic disk, induced by the merging of a minor satellite. Moreover, all the kinematical evidence
shows that, if this is the correct model, a low-latitude ($\leq 30\degr$) merging event is strongly preferred.


\section{CONCLUSIONS}
\label{s_conc}

We have analyzed a sample of $\sim$400 thick disk stars, measuring the variation of their kinematical properties
as a function of distance from the Galactic plane, from $\vert Z \vert$=1.5 to 4.5~kpc. Our results can be
summarized as follows:
\begin{itemize}
\item While the mean vertical velocity component $\overline{W}$ shows no significant deviation from zero in
the whole range, between 1.5 and 3~kpc, we find a net radial motion of about 20~km~s$^{-1}$ directed toward the
Galactic anticenter. Other authors have recently found evidence for a similar behavior, proposing an inward
motion of the LSR, although our mean velocity is larger than their proposed value by a factor of two. However, we
find that $\overline{U}$ changes sign for $\vert Z \vert\geq$3~kpc, and that a radial motion of the LSR, although
not excluded, cannot alone explain this behavior.
\item The mean rotational velocity of the thick disk decreases with distance from the Galactic plane, as found by
many previous investigations. The linear fit of our data returns a gradient of $-30$~km~s$^{-1}$~kpc$^{-1}$, our
data points and the results of three other previous works are globally better represented by a power-law with
index 1.25, very similar to what has recently been proposed by \citet{Bond10}.
\item All the velocity dispersions steadily increase with distance from the Galactic plane, closely following a
linear relation. The gradients we found are, however, smaller than those proposed by previous works.
\item While the velocity dispersions increase with $\vert Z \vert$, the ratios
$\sigma_\mathrm{U}/\sigma_\mathrm{W}$ and $\sigma_\mathrm{U}/\sigma_\mathrm{V}$ show no significant vertical
trend. The observations thus indicate a substantial constancy with $\vert Z \vert$ of the anisotropy.
\item We find a non-negligible vertex deviation, increasing with $\vert Z \vert$ from values close to zero to
$\sim 20\degr$ at $\vert Z \vert$=3.5~kpc. This is consistent with previous investigations, which found a very
small vertex deviation of old stellar population close to the Galactic plane.
\item The tilt angle steadily increases with distance from the Galactic plane. As expected, the orientation of
the velocity ellipsoid in the $U$-$W$ plane results, at any $\vert Z \vert$, intermediate between alignment with
the cylindrical and spherical coordinate systems. According to calculations by \citet{Bienayme09b}, the tilt
angle at $\vert Z \vert$=2~kpc coincides with the expectation of MOND, although the extension of their models to
higher Galactic heights is required to perform a conclusive test of the underlying gravitational law.
\item The vertical trend of many kinematical quantities show deviations from linearity between 2.5 and 3.5~kpc.
The origin of these features is unknown, but it could indicate the presence of a sub-structure at this
Galactic height, such as a comoving group of stars.
\item The results are fully consistent with the model of thick disk formation through dynamical heating of
a pre-existing Galactic disk. If this is the correct scenario, a low inclination angle of the merging event
is strongly preferred. However, not all the models proposed so far could be tested by our observations, and more
simulations are required to obtain a detailed comparison able to discriminate between them.
\end{itemize}


\acknowledgments
C.M.B. and R.A.M. acknowledge support from the Chilean Centro de Astrof\'isica FONDAP No. 15010003, and the
Chilean Centro de Excelencia en Astrof\'isica y Tecnolog\'ias Afines (CATA) BASAL PFB/06. C.M.B. also thanks
F. Mauro for useful discussions. All authors acknowledge partial support from the Yale University/Universidad
de Chile collaboration. The SPM3 catalog was funded in part by grants from the US National Science Foundation,
Yale University and the Universidad Nacional de San Juan, Argentina. We warmly thank W.~F. van Altena,
V.~I. Korchagin, T.~M. Girard, and D.~I. Casetti-Dinescu for their help and suggestions.

{\it Facilities:} \facility{Du Pont (ECHELLE)}, \facility{Magellan:Clay (MIKE)},
\facility{Euler1.2m (CORALIE)}, \facility{Max Plank:2.2m (FEROS)}


\end{document}